\documentclass[notitlepage]{article}
\usepackage[T1]{fontenc}
\usepackage[american]{babel}
\usepackage{amssymb}
\usepackage[dvips]{graphicx}
\usepackage{color}
\usepackage{vmargin}

\begin{document}

\noindent \textbf{''TRUE\ TRANSFORMATIONS\ RELATIVITY''\ and\ ELECTRODYNAMICS%
}\bigskip \medskip

\noindent \qquad \qquad \textbf{Tomislav Ivezi\'{c}\\{\textit{Ru{%
\mbox
 {\it{d}\hspace{-.15em}\rule[1.25ex]{.2em}{.04ex}\hspace{-.05em}}}er Bo\v
{s}kovi\'{c} Institute, P.O.B. 180 , }10002 Zagreb, Croatia}}\\
\noindent \hspace*{13mm} ivezic@rudjer.irb.hr 
\bigskip

\noindent \qquad \qquad 
{\small \ \medskip }

\noindent Different approaches to special relativity (SR) are discussed. The
first approach is an invariant approach, which we call the ''true
transformations (TT)\ relativity.'' In this approach a physical quantity in
the four-dimensional spacetime is mathematically represented either by a
true tensor (when no basis has been introduced) or equivalently by a
coordinate-based geometric quantity comprising both components and a basis
(when some basis has been introduced). This invariant approach is compared
with the usual covariant approach, which mainly deals with the basis
components of tensors in a specific, i.e., Einstein's coordinatization of
the chosen inertial frame of reference. The third approach is the usual
noncovariant approach to SR in which some quantities are not tensor
quantities, but rather quantities from ''3+1'' space and time, e.g., the
synchronously determined spatial length. This formulation is called the
''apparent transformations (AT)\ relativity.'' It is shown that the
principal difference between these approaches arises from the difference in
the concept of sameness of a physical quantity for different observers. This
difference is investigated considering the spacetime length in the ''TT
relativity'' and spatial and temporal distances in the ''AT relativity.'' It
is also found that the usual transformations of the three-vectors
(3-vectors) of the electric and magnetic fields $\mathbf{E}$ and $\mathbf{B}$
are the AT. Furthermore it is proved that the Maxwell equations with the
electromagnetic field tensor $F^{ab}$ and the usual Maxwell equations with $%
\mathbf{E}$ and $\mathbf{B}$ are not equivalent, and that the Maxwell
equations with $\mathbf{E}$ and $\mathbf{B}$ do not remain unchanged in form
when the Lorentz transformations of the ordinary derivative operators and
the AT of $\mathbf{E}$ and $\mathbf{B}$ are used. The Maxwell equations with 
$F^{ab}$ are written in terms of the 4-vectors of the electric $E^{a}$ and
magnetic $B^{a}$ fields. The covariant Majorana electromagnetic field
4-vector $\Psi ^{a}$ is constructed by means of 4-vectors $E^{a}$ and $B^{a}$
and the covariant Majorana formulation of electrodynamics is presented. A
Dirac like relativistic wave equation for the free photon is
obtained.\bigskip \medskip 

\noindent PACS number(s): 03.30.+p \bigskip \medskip

\noindent \textbf{1. INTRODUCTION} \bigskip

\noindent At present there are two formulations of the classical
electrodynamics. The first one is the manifestly covariant formulation,
which deals with the component form, in Einstein's coordinatization, of the
tensor quantities and tensor equations in the four-dimensional (4D)
spacetime, and where the electromagnetic field tensor $F^{\alpha \beta }$
(the component form; for the notation see the next section) contains all the
information about the electromagnetic field. The second one is the
noncovariant formulation dealing with the three-vectors (3-vectors), the
electric field $\mathbf{E}$ and the magnetic field $\mathbf{B}$, and with
equations containing them. The whole latter formulation is given in ''3+1''
space and time and was constructed by Maxwell before the appearance of
Einstein's theory of relativity.$^{(1)}$

In the recent papers$^{(2,3)}$ I have presented an alternative covariant
formulation of vacuum electrodynamics with the electric and magnetic
4-vectors $E^{\alpha }$ and $B^{\alpha }$ (also the component form), which
is equivalent to the usual covariant formulation with $F^{\alpha \beta }$.
The covariant formulation with $F^{\alpha \beta }$ and the usual formulation
with $\mathbf{E}$ and $\mathbf{B}$ are generally considered to be
equivalent. It is shown in Refs. 2 and 3 that the equivalence between
covariant formulations (either the usual one with $F^{\alpha \beta }$, or
equivalently the alternative one with $E^{\alpha }$ and $B^{\alpha }$) and
the usual noncovariant formulation does not exist.

In Ref. 4 Rohrlich introduced the notions of the true transformations (TT)
and the apparent transformations (AT) of physical quantities and emphasized
the role of sameness of a physical quantity for different observers. This
concept of sameness is also considered in the same sense by Gamba.$^{(5)}$
Their formulations also refer to the component form, in Einstein's
coordinatization, of the tensor quantities and tensor equations.

In this paper we explore a formulation of special relativity (SR) that is
borrowed from general relativity. This is the formulation in which all
physical quantities (in the case when no basis has been introduced) are
described by \emph{true tensor fields}, that are defined on the 4D
spacetime, and that satisfy \emph{true tensor equations }representing
physical laws. When the coordinate system has been introduced the physical
quantities are mathematically represented by the coordinate-based geometric
quantities (CBGQs) that satisfy the coordinate-based geometric equations
(CBGEs). \emph{The CBGQs contain both the components and the basis one-forms
and vectors} of the chosen inertial frame of reference (IFR). The symmetry
transformations for the metric $g_{ab}$, i.e., the isometries, leave the
pseudo-Euclidean geometry of 4D spacetime of SR unchanged. At the same time
they do not change the true tensor quantities, or equivalently the CBGQs, in
physical equations. Thus \emph{isometries are} what Rohrlich$^{(4)}$ calls 
\emph{the TT}. The formulation of SR that deals with true tensor quantities
and the TT is called the ''TT relativity.'' In Sec. 2 we give some specific
examples of the TT, i.e., of the isometries. They are the covariant 4D
Lorentz transformation $L^{a}{}_{b}$ (\ref{fah}) (see also Refs. 3, 6 and
7), or its representations in different coordinatizations, $L^{\mu ^{\prime
}}{}_{\nu ,e}$ (\ref{lorus}) and $L^{\mu ^{\prime }}{}_{\nu ,r}$ (\ref{elr}%
), and also the transformations $T_{\;\nu }^{\mu }$ (\ref{lamb}) that
connect the Einstein coordinatization with another coordinatization of the
considered IFR. In the ''TT relativity'' different coordinatizations of an
IFR are allowed and they are all equivalent in the description of physical
phenomena. Particularly two very different coordinatizations, the Einstein
(''e'')$^{(1)}$ and ''radio'' (''r'')$^{(8)}$ coordinatization, are exposed
in Sec. 2 and exploited throughout the paper.

As explained in Sec. 2 the ''TT relativity'' approach differs from the usual
covariant approach as well. \emph{In the usual covariant approach one does
not deal with the true tensors but with the basis components of tensors
(mainly in the ''e'' coordinatization) and with the equations of physics
written out in the component form.} (In the ''e'' coordinatization the
Einstein synchronization$^{(1)}$ of distant clocks and cartesian space
coordinates $x^{i}$ are used in the chosen IFR.)

In contrast to the TT \emph{the AT} are not the transformations of spacetime
tensors and they do not refer to the same quantity. Thus they are not
isometries and they \emph{refer exclusively to the component form of tensor
quantities and in that form they transform only some components of the whole
tensor quantity.} In fact, depending on the used AT, only a part of a 4D
tensor quantity is transformed by the AT. Such a part of a 4D quantity, when
considered in different IFRs (or in different coordinatizations of some IFR)
corresponds to different quantities in 4D spacetime. Some examples of\ the
AT are: the AT of the synchronously defined spatial length,$^{(1)}$ i.e.,
the Lorentz ''contraction''$^{(4,5,3,7,9)}$ and the AT of the temporal
distance, i.e., the conventional ''dilatation'' of time that is introduced
in Ref. 1 and considered in Refs. 7 and 9, and as discussed in Refs. 2 and
3, the AT of the electric and magnetic 3-vectors $\mathbf{E}$ and $\mathbf{B}
$ (the conventional transformations of $\mathbf{E}$ and $\mathbf{B,}$ see,
e.g., Ref. 10 Sec.11.10). The formulation of SR which uses the AT we call
the ''apparent transformations (AT) relativity.'' An example of such
formulation is Einstein's formulation of SR which is based on his two
postulates and which deals with all the mentioned AT.

The differences between the TT and the AT are examined considering some
specific examples. First the spacetime lengths, corresponding in ''3+1''
picture to a moving rod and to a moving clock, are considered in the ''TT
relativity'' in Secs. 3.1 and 3.2. Furthermore the spatial and temporal
distances for the same examples are examined in the ''AT relativity'' in
Secs. 4.1 and 4.2. The comparison with the experiments on the Lorentz
contraction and the time dilatation is performed in Ref. 9 and it shows that
all experiments can be qualitatively and quantitatively explained by the
''TT relativity,'' while some experiments cannot be adequately explained by
the ''AT relativity.''

The second part of this paper is an electrodynamic part in which it is
proved that the conventional transformations of $\mathbf{E}$ and $\mathbf{B}$
are the AT, Secs. 5.1-5.3, that the usual Maxwell equations with the
3-vectors $\mathbf{E}$ and $\mathbf{B}$ are not equivalent with the tensor
Maxwell's equations with $F^{ab}$, and that the Maxwell equations with $%
\mathbf{E}$ and $\mathbf{B}$ change their form when the Lorentz
transformations of the ordinary derivative operators and the usual AT of $%
\mathbf{E}$ and $\mathbf{B}$ are used, Sec. 5.3. The tensor Maxwell
equations with $E^{a}$ and $B^{a}$ were presented and shown to be completely
equivalent to the tensor Maxwell's equations with $F^{ab}$ in Sec. 6 (see
also$^{(2,3)}$ for the component form)$.$ The expression for the Lorentz
force in terms of $E^{a}$ and $B^{a},$ and also the equation of motion of a
charge $q$ moving in the electromagnetic field $E^{a}$ and $B^{a}$ are
presented in Sec. 6.1. The comparison of this invariant approach with $E^{a}$
and $B^{a}$ and the usual noncovariant approach with the 3-vectors $\mathbf{E%
}$ and $\mathbf{B}$ is possible in the ''e'' coordinatization, and it is
done in Sec. 6.2. Then in Sec. 6.3 we construct the covariant Majorana
electromagnetic field four-vector $\Psi ^{a}$ by means of four-vectors $%
E^{a} $ and $B^{a}$ and present the covariant Majorana formulation of
electrodynamics. An important advantage of the approach with $E^{a},$ $B^{a}$
and $\Psi ^{a}$ is that this formulation of electrodynamics does not make
use of the intermediate electromagnetic 4-potential $A^{\mu }$, and thus
dispenses with the need for gauge conditions. Using Majorana formulation we
find a Dirac like relativistic wave equation for the free photon in Sec. 6.3
(see also$^{(3)}$ ). In Sec. 7 we present summary and conclusions.\bigskip
\medskip

\noindent \textbf{2. A\ GENERAL\ DISCUSSION\ ON\ THE\ ''TT\ RELATIVITY''}\
\bigskip

The principal difference between the ''TT relativity,'' the usual covariant
formulation and the ''AT relativity'' stems from the difference in the
concept of \emph{sameness} of a physical system, i.e., of a physical
quantity, for different observers. This concept actually determines the
difference in what is to be understood as a relativistic theory.

Thus in the ''TT relativity'' a physical quantity is mathematically
represented either by a true tensor or by a CBGQ comprising both components
and a basis. (Speaking in mathematical language a tensor of type (k,l) is
defined as a linear function of k one-forms and l vectors (in old names, k
covariant vectors and l contravariant vectors) into the real numbers, see,
e.g., Refs. 11, 12 and 13. If a coordinate system is chosen in some IFR
then, in general, any tensor quantity can be reconstructed from its
components and from the basis vectors and basis 1-forms of that frame, i.e.,
it can be written in a coordinate-based geometric language.$^{(13)}$) \emph{%
The CBGQs representing some 4D physical quantity in different relatively
moving inertial frames of reference (IFRs), or in different
coordinatizations of the chosen IFR, are all mathematically equal. }Thus
they are really \emph{the same quantity }for different observers, or in
different coordinatizations. We suppose that in the ''TT relativity'' such
4D tensor quantities are well-defined not only mathematically but also \emph{%
experimentally}, as measurable quantities with real physical meaning. \emph{%
The complete and well-defined measurement from the ''TT relativity''
viewpoint is such measurement in which all parts of some 4D quantity are
measured.}

\emph{In the usual covariant approach (including Refs. 4 and 5) the basis
components of a true tensor, or equivalently of a CBGQ, that are determined
in different IFRs (or in different coordinatizations), are considered to be
the same quantity for different observers.} It is true that these components
refer to the same tensor quantity but they cannot be equal since the bases
are not included. Mathematically speaking \emph{the concept of a tensor in
the usual covariant approach is defined entirely in terms of the
transformation properties of its components relative to some coordinate
system. }

In the ''TT relativity,'' as we already said, CBGQs (comprising both
components and a basis), which represent some 4D physical quantity for
different observers, are all mathematically equal. Therefore \emph{the ''TT
relativity'' }approach (which deals with the true tensors or with the CBGQs) 
\emph{is an invariant approach} to SR in contrast to the usual \emph{%
covariant approach} (which deals with the components of tensors). Obviously
one can use different possible coordinatizations in the ''TT relativity''
approach. We shall explicitly use two very different coordinatizations, the
''e'' and ''r'' coordinatizations, see below.

All this can be illustrated by a simple example, i.e., considering the
distance 4-vector (the $(1,0)$ tensor) $l_{AB}^{a}=x_{B}^{a}-x_{A}^{a}$
between two events $A$ and $B$ (with the position 4-vectors $x_{A}^{a}$ and $%
x_{B}^{a}$). It can be equivalently represented in the coordinate-based
geometric language in different bases, $\left\{ e_{\mu }\right\} $ and $%
\left\{ r_{\mu }\right\} $ in an IFR $S,$ and $\left\{ e_{\mu ^{\prime
}}\right\} $ and $\left\{ r_{\mu ^{\prime }}\right\} $ in a relatively
moving IFR $S^{\prime },$ as $l_{AB}^{a}=l_{e}^{\mu }e_{\mu }=l_{r}^{\mu
}r_{\mu }=l_{e}^{\nu ^{\prime }}e_{\nu ^{\prime }}=l_{r}^{\mu ^{\prime
}}r_{\mu ^{\prime }},$ where, e.g., $e_{\mu }$ are the basis 4-vectors, $%
e_{0}=(1,0,0,0)$ and so on, and $l_{e}^{\mu }$ are the basis components when
the Einstein coordinatization is chosen in some IFR $S.$ The subscript $%
^{\prime }e^{\prime }$ stands for the Einstein coordinatization and the
subscript $^{\prime }r^{\prime }$ for the ''r'' coordinatization, see below.
The decompositions $l_{e}^{\mu }e_{\mu }$ and $l_{r}^{\mu }r_{\mu }$ (in an
IFR $S,$ and in the ''e'' and ''r'' coordinatizations respectively) and $%
l_{e}^{\nu ^{\prime }}e_{\nu ^{\prime }}$ and $l_{r}^{\mu ^{\prime }}r_{\mu
^{\prime }}$ (in a relatively moving IFR $S^{\prime }$, and in the ''e'' and
''r'' coordinatizations respectively) of the true tensor $l_{AB}^{a}$ are
all mathematically \emph{equal} quantities and thus they are really the same
quantity considered in different relatively moving IFRs and in different
coordinatizations. This is the treatment of the distance 4-vector in the
''TT relativity.'' On the other hand the usual covariant approach does not
consider the whole tensor quantity, the distance 4-vector $l_{AB}^{a},$ but
only the basis components, $l_{e}^{\mu }$ and $l_{e}^{\nu ^{\prime }},$ in
the ''e'' coordinatization. Note that, in contrast to the above equalities
for the CBGQs, the sets of components, e.g., $l_{e}^{\mu }$ and $l_{e}^{\nu
^{\prime }},$ taken alone, are not equal, $l_{e}^{\mu }\neq l_{e}^{\nu
^{\prime }},$ and thus they are not the same quantity from the ''TT
relativity'' viewpoint. From the mathematical point of view the components
of, e.g., a $(1,0)$ tensor are its values (real numbers) when the basis
one-form, for example, $e^{\alpha },$ is its argument (see, e.g., Ref. 12).
Thus, for example, $l_{AB}^{a}(e^{\alpha })=l_{e}^{\mu }e_{\mu }(e^{\alpha
})=l_{e}^{\alpha }$ (where $e^{\alpha }$ is the basis one-form in an IFR $S$
and in the ''e'' coordinatization), while $l_{AB}^{a}(e^{\alpha ^{\prime
}})=l_{e}^{\mu ^{\prime }}e_{\mu ^{\prime }}(e^{\alpha ^{\prime
}})=l_{e}^{\alpha ^{\prime }}$ (where $e^{\alpha ^{\prime }}$ is the basis
one-form in $S^{\prime }$ and in the ''e'' coordinatization). Obviously $%
l_{e}^{\alpha }$ and $l_{e}^{\alpha ^{\prime }}$ are not the same real
numbers since the basis one-forms $e^{\alpha }$ and $e^{\alpha ^{\prime }}$
are different bases.

In the above discussion and henceforward I adopt the following convention
with regard to indices. Repeated indices imply summation. Latin indices $%
a,b,c,d,...$ are to be read according to the abstract index notation, Ref.
11, Sec.2.4.; they ''...should be viewed as reminders of the number and type
of variables the tensor acts on, \emph{not} as basis components.'' They
designate geometric objects in 4D spacetime. Thus, e.g., $l_{AB}^{a},$ $%
x_{A}^{a}$ are (1,0) tensors and they are defined independently of any
coordinate system. Greek indices run from 0 to 3, while latin indices $%
i,j,k,l,...$ run from 1 to 3, and they both designate the components of some
geometric object in some coordinate system, e.g., $x^{\mu }(x^{0},x^{i})$
and $x^{\mu ^{\prime }}(x^{0^{\prime }},x^{i^{\prime }})$ are two coordinate
representations of the position 4-vector $x^{a}$ in two different inertial
coordinate systems $S$ and $S^{\prime }.$ Similarly the metric tensor $%
g_{ab} $ denotes a tensor of type (0,2) (whose Riemann curvature tensor $%
R_{bcd}^{a} $ is everywhere vanishing; the spacetime of special relativity
is a flat spacetime, and this definition includes not only the IFRs but also
the accelerated frames of reference). This geometric object $g_{ab}$ is
represented in some IFR $S,$ and in the ''e'' coordinatization, by the $%
4\times 4$ diagonal matrix of components of $g_{ab}$, $g_{\mu \nu
,e}=diag(-1,1,1,1),$ and this matrix is usually called the Minkowski metric
tensor$.$

We shall also need the expression for the covariant 4D Lorentz
transformations $L^{a}{}_{b}$,$^{(6,3,7)}$ which is independent of the
chosen synchronization, i.e., coordinatization of reference frames. It is 
\begin{equation}
L^{a}{}_{b}\equiv L^{a}{}_{b}(v)=g^{a}{}_{b}-\frac{2u^{a}v_{b}}{c^{2}}+\frac{%
(u^{a}+v^{a})(u_{b}+v_{b})}{c^{2}(1+\gamma )}  \label{fah}
\end{equation}
where $u^{a}$ is the proper velocity 4-vector of a frame $S$ with respect to
itself (only $u^{0}\neq 0$, see also Ref. 6). $u^{a}$ can be written as $%
u^{a}=cn^{a},$ $n^{a}$ is the unit 4-vector along the $x^{0}$ axis of the
frame $S.$ $v^{a}$ in (\ref{fah}) is the proper velocity 4-vector of $%
S^{\prime }$ relative to $S.$ Further $u\cdot v=u^{a}v_{a}$ and $\gamma
=-u\cdot v/c^{2}.$ When we use the Einstein coordinatization then $%
L^{a}{}_{b}$ is represented by $L^{\mu ^{\prime }}{}_{\nu ,e},$ the usual
expression for pure Lorentz transformation, but with $v_{e}^{i}$ (the proper
velocity 4-vector $v_{e}^{\mu }$ is $v_{e}^{\mu }\equiv dx_{e}^{\mu }/d\tau
=(\gamma _{e}c,\gamma _{e}v_{e}^{i}),$ $d\tau \equiv dt_{e}/\gamma _{e}$ is
the scalar proper-time, and $\gamma _{e}\equiv (1-v_{e}^{2}/c^{2})^{1/2}$)
replacing the components of the ordinary velocity 3-vector $\mathbf{V.}$
Obviously, in the usual form, the Lorentz transformation connect two
coordinate representations, basis components (in the ''e'' coordinatization) 
$x_{e}^{\mu },$ $x_{e}^{\mu ^{\prime }}$ of a given event; $x_{e}^{\mu },$ $%
x_{e}^{\mu ^{\prime }}$ refer to two relatively moving IFRs (with the
Minkowski metric tensors) $S$ and $S^{\prime },$ 
\begin{eqnarray}
x_{e}^{\mu ^{\prime }} &=&L^{\mu ^{\prime }}{}_{\nu ,e}x_{e}^{\nu },\qquad
\,L^{0^{\prime }}{}_{0,e}=\gamma _{e},\quad L^{0^{\prime
}}{}_{i,e}=L^{i^{\prime }}{}_{0,e}=-\gamma _{e}v_{e}^{i}/c  \nonumber \\
L^{i^{\prime }}{}_{j,e} &=&\delta _{j}^{i}+(\gamma
_{e}-1)v_{e}^{i}v_{je}/v_{e}^{2}  \label{lorus}
\end{eqnarray}
Since $g_{\mu \nu ,e}$ is a diagonal matrix the space $x_{e}^{i}$ and time $%
t_{e}$ $(x_{e}^{0}\equiv ct_{e})$ parts of $x_{e}^{\mu }$ do have their
usual meaning.

The geometry of the spacetime is generally defined by the metric tensor $%
g_{ab},$ which can be expand in a coordinate basis in terms of its
components as $g_{ab}=g_{\mu \nu }dx^{\mu }\otimes dx^{\nu },$ and where $%
dx^{\mu }\otimes dx^{\nu }$ is an outer product of the basis 1-forms.
However the 'old-fashioned' notation $ds^{2}$ is often used in place of $%
g_{ab}$ to represent the metric tensor$,$ and one writes the relation for $%
g_{ab}$ as $ds^{2}=g_{\mu \nu }dx^{\mu }dx^{\nu }$ omitting the outer
product sign. The metric is then understood as representing the invariant
infinitesimal squared spacetime distance of two neighboring points. Here we
follow the same practice of using $ds^{2}.$ In the ''e'' coordinatization
the geometrical quantity $ds^{2}$ can be written in terms of its
representation $ds_{e}^{2},$ with the separated spatial and temporal parts, $%
ds^{2}=ds_{e}^{2}=(dx_{e}^{i}dx_{ie})-(dx_{e}^{0})^{2}$, and the same
happens with the spacetime length $l,$ the invariant spacetime length (the
Lorentz scalar) between two points (events) in 4D spacetime. $l$ is defined
as 
\begin{equation}
l=(g_{ab}l^{a}l^{b})^{1/2}  \label{elspat}
\end{equation}
where $l^{a}(l^{b})$ is the distance 4-vector between two events $A$ and $B$%
, $l^{a}=l_{AB}^{a}=x_{B}^{a}-x_{A}^{a}$, and it holds that $%
l^{2}=l_{e}^{2}=(l_{e}^{i}l_{ie})-(l_{e}^{0})^{2}$. Such separation remains
valid in other inertial coordinate systems with the Minkowski metric tensor,
and in $S^{\prime }$ one finds $l^{2}=l_{e}^{\prime 2}=(l_{e}^{i^{\prime
}}l_{i^{\prime }e})-(l_{e}^{0^{\prime }})^{2},$ where $l_{e}^{\mu ^{\prime
}} $ in $S^{\prime }$ is connected with $l_{e}^{\mu }$ in $S$ by the Lorentz
transformation $L^{\mu ^{\prime }}{}_{\nu ,e}$ (\ref{lorus}).

As shown in Refs. 3 and 7 we can also choose another coordinatization, the
''everyday'' or ''radio'' (''r'') coordinatization,$^{(8)}$ which differs
from the ''e'' coordinatization by the different procedure for the
synchronization of distant clocks. Different synchronizations are determined
by the parameter $\varepsilon $ in the relation $t_{2}=t_{1}+\varepsilon
(t_{3}-t_{1})$, where $t_{1}$ and $t_{3}$ are the times of departure and
arrival, respectively, of the light signal, read by the clock at $A$, and $%
t_{2}$ is the time of reflection at $B$, read by the clock at $B$, that has
to be synchronized with the clock at $A$. In Einstein's synchronization
convention $\varepsilon =1/2.$ In the ''r'' synchronization $\varepsilon =0$
and thus, in contrast to the ''e'' synchronization, there is an absolute
simultaneity. As explained in Ref. 8: ''For if we turn on the radio and set
our clock by the standard announcement ''...at the sound of the last tone,
it will be 12 o'clock'', then we have synchronized our clock with the studio
clock in a manner that corresponds to taking $\varepsilon =0$ in $%
t_{2}=t_{1}+\varepsilon (t_{3}-t_{1}).$'' The ''r'' synchronization$^{(8)}$
is an assymetric synchronization which leads to an assymetry in the
coordinate, one-way, speed of light. However from the physical point of view
the ''r'' coordinatization is completely equivalent to the ''e''
coordinatization. This also holds for all other permissible
coordinatizations. Such situation really happens in the ''TT relativity''
since the ''TT relativity'' deals with true tensors and the true tensor
equations\emph{,} or equivalently with the CBGQs and CBGEs. Thus the ''TT
relativity'' deals on the same footing with all possible coordinatizations
of the chosen IFR. As a consequence \emph{the second Einstein postulate
referred to the constancy of the coordinate velocity of light, in general,
does not hold in the ''TT relativity.''} Namely, only in Einstein's
coordinatization the coordinate, one-way, speed of light is isotropic and
constant.

The basis vectors in the ''r'' cordinatization are constructed as in Refs. 8
and 3, and here we expose this construction once again for the sake of
clearness of the whole exposition. The temporal basis vector $e_{0}$ is the
unit vector directed along the world line of the clock at the origin. The
spatial basis vectors by definition connect \emph{simultaneous} events, the
event ''clock at rest at the origin reads 0 time'' with the event ''clock at
rest at unit distance from the origin reads 0 time,'' and thus they are
synchronization-dependent. The spatial basis vector $e_{i}$ connects two
above mentioned simultaneous events when Einstein's synchronization ($%
\varepsilon =1/2$) of distant clocks is used. The temporal basis vector $%
r_{0}$ is the same as $e_{0}.$ The spatial basis vector $r_{i}$ connects two
above mentioned simultaneous events when ''radio'' clock synchronization ($%
\varepsilon =0$) of distant clocks is used. The spatial basis vectors, e.g., 
$r_{1},r_{1}^{\prime },r_{1}^{\prime \prime }..$ are parallel and directed
along an (observer-independent) light line. Hence, two events that are
everyday (''r'') simultaneous in $S$ are also ''r'' simultaneous for all
other IFRs.

The connection between the basis vectors in the ''r'' and ''e''
coordinatizations is given as $r_{0}=e_{0},\;r_{i}=e_{0}+e_{i}$. The metric
tensor $g_{ab}$ becomes $g_{ab}=g_{\mu \nu ,r}dx_{r}^{\mu }\otimes
dx_{r}^{\nu }$ in the coordinate-based geometric language and in the ''r''
coordinatization, where the basis components of the metric tensor are $%
g_{00,r}=g_{0i,r}=g_{i0,r}=g_{ij,r}(i\neq j)=-1,g_{ii,r}=0$. $dx_{r}^{\mu },$
$dx_{r}^{\nu }$ are the basis 1-forms in the ''r'' coordinatization and in $%
S,$ and $dx_{r}^{\mu }\otimes dx_{r}^{\nu }$ is an outer product of the
basis 1-forms, i.e., it is the basis for (0,2) tensors.

The transformation matrix $T_{\;\nu ,r}^{\mu }$ which transforms the ''e''
coordinatization to the ''r'' coordinatization is given as $T_{\;\mu
,r}^{\mu }=-T_{\;i,r}^{0}=1,$ and all other elements of $T_{\;\nu ,r}^{\mu }$
are $=0$. $T_{\;\nu ,r}^{\mu }$ is obtained from Logunov's$^{(14)}$
expression for the transformation matrix $\lambda _{\nu }^{\overline{\mu }}$
connecting (in his interpretation) a physicaly measurable tensor with the
coordinate one. Thus in the approach$^{(14)}$ there are physical and
coordinate quantities for the same coordinatization of the considered IFR.
However, in our interpretation, his ''physicaly measurable tensor''
corresponds to the tensor written in the Einstein coordinatization of a
given IFR, and the coordinate one corresponds to some arbitrary
coordinatization of the same IFR. Thence, his matrix $\lambda _{\nu }^{%
\overline{\mu }}$ can be interpreted as the transformation matrix between
some arbitrary coordinatization and the ''e'' coordinatization. The elements
of $\lambda _{\nu }^{\overline{\mu }}$, which are different from zero, are $%
\lambda _{0}^{\overline{0}}=(-g_{00})^{1/2},$ $\lambda _{i}^{\overline{0}%
}=(-g_{0i})(-g_{00})^{-1/2},\quad \lambda _{i}^{\overline{i}}=\left[
g_{ii}-(g_{0i})^{2}/g_{00}\right] ^{1/2}.$ We actually need the inverse
transformation $(\lambda _{\nu }^{\overline{\mu }})^{-1}$ (it will be
denoted as $T_{\;\nu }^{\mu }$ to preserve the similarity with the notation
from$^{(3)}$). Then the elements (that are different from zero) of the
matrix $T_{\;\nu }^{\mu },$ which transforms the ''e'' coordinatization to
the coordinatization determined by the basis components $g_{\mu \nu }$ of
the metric tensor $g_{ab},$ are 
\begin{eqnarray}
T_{\;0}^{0} &=&(-g_{00})^{-1/2},\quad
T_{\;i}^{0}=(g_{0i})(-g_{00})^{-1}\left[ g_{ii}-(g_{0i})^{2}/g_{00}\right]
^{-1/2}  \nonumber \\
T_{\;i}^{i} &=&\left[ g_{ii}-(g_{0i})^{2}/g_{00}\right] ^{-1/2}  \label{lamb}
\end{eqnarray}
The transformation matrix $T_{\;\nu ,r}^{\mu }$ is then easily obtained from 
$T_{\;\nu }^{\mu }$ (\ref{lamb}) and the known $g_{\mu \nu ,r}$. We note
that in SR, i.e., in the theory of flat spacetime, any specific $g_{\mu \nu
} $ (for the specific coordinatization) can be transformed to the Minkowski
matrix $g_{\mu \nu ,e}$. It can be accomplished by means of the matrix $%
(T_{\;\nu }^{\mu })^{-1}$; for example, $g_{\mu \nu ,r}$ is transformed by
the matrix $(T_{\;\nu ,r}^{\mu })^{-1}$ to $g_{\mu \nu ,e}.$

For the sake of completeness we also quote the Lorentz transformation $%
L^{\mu ^{\prime }}\,_{\nu ,r}$ in the ''r'' coordinatization. It can be
easily found from $L^{a}{}_{b}$ (\ref{fah}) and the known $g_{\mu \nu ,r},$
and the elements that are different from zero are 
\begin{eqnarray}
x_{r}^{\prime \mu } &=&L^{\mu ^{\prime }}{}_{\nu ,r}x_{r}^{\nu },\quad
L^{0^{\prime }}{}_{0,r}=K,\quad L^{0^{\prime }}{}_{2,r}=L^{0^{\prime
}}{}_{3,r}=K-1  \nonumber \\
L^{1^{\prime }}{}_{0,r} &=&L^{1^{\prime }}{}_{2,r}=L^{1^{\prime
}}{}_{3,r}=(-\beta _{r}/K),L^{1^{\prime }}{}_{1,r}=1/K,\quad L^{2^{\prime
}}{}_{2,r}=L^{3^{\prime }}{}_{3,r}=1  \label{elr}
\end{eqnarray}
where $K=(1+2\beta _{r})^{1/2},$ and $\beta _{r}=dx_{r}^{1}/dx_{r}^{0}$ is
the velocity of the frame $S^{\prime }$ as measured by the frame $S$, $\beta
_{r}=\beta _{e}/(1-\beta _{e})$ and it ranges as $-1/2\prec \beta _{r}\prec
\infty .$ Since $g_{\mu \nu ,r},$ in contrast to $g_{\mu \nu ,e},$ is not a
diagonal matrix, then in $ds_{r}^{2}$ the spatial and temporal parts are not
separated, and the same holds for the spacetime length $l,$ see Ref. 3 for
the results in 2D spacetime. Expressing $dx_{r}^{\mu },$ or $l_{r}^{\mu }$,
in terms of $dx_{e}^{\mu },$ or $l_{e}^{\mu },$ one finds that $%
ds^{2}=ds_{r}^{2}=ds_{e}^{2},$ and also, $l^{2}=l_{r}^{2}=l_{e}^{2},$ as it
must be. It can be easily proved$^{(8)}$ that the ''r'' synchronization is
an assymetric synchronization which leads to an assymetry in the measured
''one-way'' velocity of light (for one direction $c_{r}^{+}=\infty $ whereas
in the opposite direction $c_{r}^{-}=-c/2$). The round trip velocity,
however, does not depend on the chosen synchronization procedure, and it is $%
\equiv c.$ The coordinate system in which $g_{0i}=0$ at every point in 4D
spacetime is called time-orthogonal since in it the time axis is everywhere
orthogonal to the spatial coordinate curves. This happens in the cases when
in some IFR the Einstein synchronization is chosen together with, e.g.,
cartesian, or polar, or spherical, etc., spatial coordinates. However it is
not the case when the ''r'' synchronization is chosen. It has to be noted
that although in the Einstein coordinatization the space and time components
of the position 4-vector do have their usual meaning, i.e., as in the
prerelativistic physics, and in $ds_{e}^{2}$ the spatial and temporal parts
are separated, it does not mean that the ''e'' coordinatization does have
some advantage relative to other coordinatizations and that the quantities
in the ''e'' coordinatization are more physical (for the recent discussion
of the conventionality of synchronization see Ref. 15 and references
therein).

The laws of physics in SR obey the principle of \emph{special covariance},
see, e.g., Ref. 11. It says that any physically possible set of measurements
(on physical fields) obtained by a family, $O,$ of observers also is a
physically possible set of measurements for another family, $O^{\prime }$,
of observers. The observers $O^{\prime }$ are obtained by ''acting'' on $O$
with a symmetry transformation for the metric $g_{ab}.$ Such transformation
is called an isometry and it does not change $g_{ab}$; if we denote an
isometry as $\Phi ^{*}$ then $(\Phi ^{*}g)_{ab}=g_{ab}.$ Thus an isometry
leaves the pseudo-Euclidean geometry of 4D spacetime of SR unchanged. An
example of isometry is the covariant 4D Lorentz transformation $L^{a}{}_{b}$
(\ref{fah}). Note that the whole formulation is given without introducing or
making any reference to coordinate systems. In many treatments, it is
assumed that a coordinate system has been chosen and the tensor equations
expressing physical laws have been written out in component form using the
coordinate basis. However, the component form of tensor equations is not, in
general, equivalent to the true tensor equations, but only in the case when
one coordinatization is always used, i.e., as customarily assumed, the ''e''
coordinatization. When different coordinatizations of an IFR are allowed
then the true tensor equations cannot be equivalently represented by
component form of equations (as usually argued, e.g., Ref. 11), but they are
equivalent to the CBGEs, which contain both the components and the basis
one-forms and vectors of the chosen IFR. It is also stated in Ref. 11 that
if the coordinate basis is introduced the special covariance can be viewed
as expressing the invariance of the component form of equations under
isometries. However, as emphasized above, this invariance of the component
form of equations under isometries, in fact, refers to the case when only 
\emph{one} coordinatization, the ''e'' coordinatization, is \emph{always}
used. In the more general case, that is considered in this paper, when
different coordinatizations are permissible, i.e., in the ''TT relativity,''
the special covariance refers either to true tensor equations or to the
CBGEs. Thus \emph{the TT are nothing else but - the isometries, and the ''TT
relativity'' is the formulation of SR in which physical quantities are
described by true tensor fields (or by the CBGQs) that satisfy true tensor
equations (the CBGEs) representing physical laws. }

When no coordinate system is introduced the isometries act directly on
tensors and that is an 'active' point of view on transformations. However
when coordinate systems are introduced an isometry can be viewed as inducing
the coordinate transformation, e.g., $x_{e}^{\mu }\rightarrow x_{e}^{\mu
^{\prime }},$ but leaving an event $p$ and all tensors at $p$ unchanged.
This is a ''passive'' point of view on transformations. When we use the
coordinate-based geometric language then, in fact, we deal with the
''passive'' viewpoint on transformations. In practice these two viewpoints
are equivalent. Thence we can consider and write that, e.g., the distance
4-vector, the (1,0) tensor, $l_{AB}^{a}=x_{B}^{a}-x_{A}^{a}$ is equal to the
CBGQs $l_{AB}^{a}=l_{e}^{\mu }e_{\mu }=l_{r}^{\mu }r_{\mu }=l_{e}^{\nu
^{\prime }}e_{\nu ^{\prime }}=l_{r}^{\mu ^{\prime }}r_{\mu ^{\prime }},$ and
that all of them represent the same physical quantity, which remains
unchanged under the TT, i.e., under the isometries. (For the definitions of
the special covariance, the isometries and the 'active' and 'passive'
viewpoints see Ref. 11 Secs. 2 and 4 and appendix C, and for an interesting
treatment of Einstein's view of spacetime and the modern view see Ref. 16.)
Thus, in such an interpretation, the isometries do not change the tensor
quantities, or equivalently the CBGQs, in physical equations. This means
that, in contrast to \emph{the invariance of form }of the equations with
components under isometries, i.e.,\emph{\ the covariance }of such equations,
the true tensor equations, or the CBGEs, \emph{are invariant equations}
under isometries. This will be explicitly shown for different forms of the
Maxwell equations in Sec. 5.3. When the coordinate basis is introduced then,
for example, the isometry $L^{a}{}_{b}$ (\ref{fah}) will be expressed as the
coordinate Lorentz transformation $L^{\mu ^{\prime }}{}_{\nu ,e}$ (\ref
{lorus}) in the ''e'' coordinatization or $L^{\mu ^{\prime }}{}_{\nu ,r}$ (%
\ref{elr}) in the ''r'' coordinatization. The basis components will be
transformed, e.g., by $L^{\mu ^{\prime }}{}_{\nu ,e}$ while the basis
vectors $e_{\mu }$ by the inverse transformation $(L^{\mu ^{\prime }}{}_{\nu
,e})^{-1}=L^{\mu }{}_{\nu ^{\prime },e}.$

It will be also shown below that the AT, in contrast to the TT, are not
isometries and, actually, \emph{the AT refer exclusively to the component
form of tensor quantities and in that form they transform only some
components of the whole tensor quantity.} In the ''AT relativity'' such
parts of a whole tensor quantity are connected by the AT and they are
considered to be the same quantity for different observers. The examples
will be given below. \emph{\bigskip \medskip }

\noindent \textbf{3. THE\ TT\ OF\ THE\ SPACETIME\ LENGTH}\ \bigskip

The whole above discussion about the differences between the three
approaches to SR will be illustrated considering some examples, the
spacetime length with its TT and the spatial and temporal distances with
their AT.\bigskip \medskip 

\noindent \textbf{3.1. The Spacetime Length for a Moving Rod }\medskip

Let us take, for simplicity, to work in 2D spacetime. Then we also take a
particular choice for the 4-vector $l_{AB}^{a}.$ In the usual ''3+1''
picture it corresponds to an object, a rod, that is at rest in an IFR $S$
and situated along the common $x_{e}^{1},x_{e}^{1^{\prime }}-$ axes. (Its
rest length is denoted as $l_{0}.$) The situation is depicted in Fig. 1, and
the same example is already considered in Refs. 3 and 7. The decomposition
of the geometric quantity $l_{AB}^{a}$ in the ''e'' coordinatization and in $%
S$ is $l_{AB}^{a}=l_{e}^{0}e_{0}+l_{e}^{1}e_{1}=0e_{0}+l_{0}e_{1},$ while in 
$S^{\prime },$ where the rod is moving, it becomes $l_{AB}^{a}=-\beta
_{e}\gamma _{e}l_{0}e_{0^{\prime }}+\gamma _{e}l_{0}e_{1^{\prime }},$ and,
as mentioned above, 
\begin{equation}
l_{AB}^{a}=0e_{0}+l_{0}e_{1}=-\beta _{e}\gamma _{e}l_{0}e_{0^{\prime
}}+\gamma _{e}l_{0}e_{1^{\prime }}  \label{trucon}
\end{equation}
$l_{AB}^{a}$ is a tensor of type (1,0) and in (\ref{trucon}) it is written
in the coordinate-based geometric language in terms of basis vectors $e_{0},$
$e_{1},$ ($e_{0^{\prime }},$ $e_{1^{\prime }}$) and the basis components $%
l_{e}^{\mu }$ ($l_{e}^{\mu ^{\prime }}$) of some IFR.

Then we can also write the decompositions of the tensor $l_{AB}^{a}$ in the
''r'' cordinatization as 
\begin{equation}
l_{AB}^{a}=-l_{0}r_{0}+l_{0}r_{1}=-Kl_{0}r_{0^{\prime }}+(1+\beta
_{r})(1/K)l_{0}r_{1^{\prime }}  \label{trur}
\end{equation}
where $K=(1+2\beta _{r})^{1/2}.$ Remark that there is $l_{r}^{0}\neq 0,$
while in the ''e'' coordinatization $l_{e}^{0}=0,$ see Fig. 1.

\emph{It can be seen from Eqs. (\ref{trucon}) and (\ref{trur}) that the
basis components }$l_{e,r}^{\mu }$\emph{\ in }$S$\emph{\ and }$l_{e,r}^{\mu
^{\prime }}$\emph{\ in }$S^{\prime },$\emph{\ when taken alone, are not the
same 4D quantity. Only the geometric quantity }$l_{AB}^{a},$\emph{\ i.e.,
the CBGQs }$l_{e}^{\mu }e_{\mu }=l_{e}^{\mu ^{\prime }}e_{\mu ^{\prime
}}=l_{r}^{\mu }r_{\mu }=l_{r}^{\mu ^{\prime }}r_{\mu ^{\prime }}$\emph{\
comprising both, components and the basis, is the same 4D quantity for
different relatively moving IFRs; Ref. 12: ''....the components tell only
part of the story. The basis contains the rest of information.'' }Thus it
holds that 
\begin{equation}
l_{AB}^{a}=l_{e}^{\mu }e_{\mu }=l_{e}^{\mu ^{\prime }}e_{\mu ^{\prime
}}=l_{r}^{\mu }r_{\mu }=l_{r}^{\mu ^{\prime }}r_{\mu ^{\prime }}
\label{elt5}
\end{equation}
These results can be clearly understood from Fig. 1 (see also Ref. 7). We
see from (\ref{trucon}) that in the ''e'' coordinatization, that is commonly
used in the ''AT relativity,'' there is a dilatation of the spatial part $%
l_{e}^{1^{\prime }}=\gamma _{e}l_{0}$ with respect to $l_{e}^{1}=l_{0}$ and
not the Lorentz contraction as predicted in the ''AT relativity.'' Similarly
if only spatial parts of $l_{r}^{\mu }$ and $l_{r}^{\mu ^{\prime }}$ are
compared then one finds the dilatation $\infty \succ l_{r}^{1^{\prime }}\geq
l_{0}$ for all $\beta _{r}$. Hovewer it is clear from the above discussion
that comparison of only spatial parts of the components of the distance
4-vector $l_{AB}^{a}$ in $S$ and $S^{\prime }$ is physically meaningless in
the ''TT relativity,'' since \emph{some components of the tensor quantity,
when they are taken alone, do not correspond to some definite 4D physical
quantity.} Also we remark that the whole tensor quantity $l_{AB}^{a}$\
comprising components and the basis is transformed by the Lorentz
transformation from $S$\ to $S^{\prime }.$ Note that if $l_{e}^{0}=0$ then $%
l_{e}^{\mu ^{\prime }}$ in any other IFR $S^{\prime }$ will contain the time
component $l_{e}^{0^{\prime }}\neq 0.$ The spacetime length for the
considered case is $l=(l_{e,r}^{\mu }l_{\mu e,r})^{1/2}=(l_{e,r}^{\mu
^{\prime }}l_{\mu ^{\prime }e,r})^{1/2}=l_{0}.$ In the ''e''
coordinatization and in $S,$ the rest frame of the rod, where the temporal
part of $l_{e}^{\mu }$ is $l_{e}^{0}=0,$ the spacetime length $l$ is a
measure of the spatial distance, i.e., of the rest spatial length of the
rod, as in the prerelativistic physics.\bigskip \medskip
\begin{figure}
\begin{center}
\includegraphics[width=8.0cm]{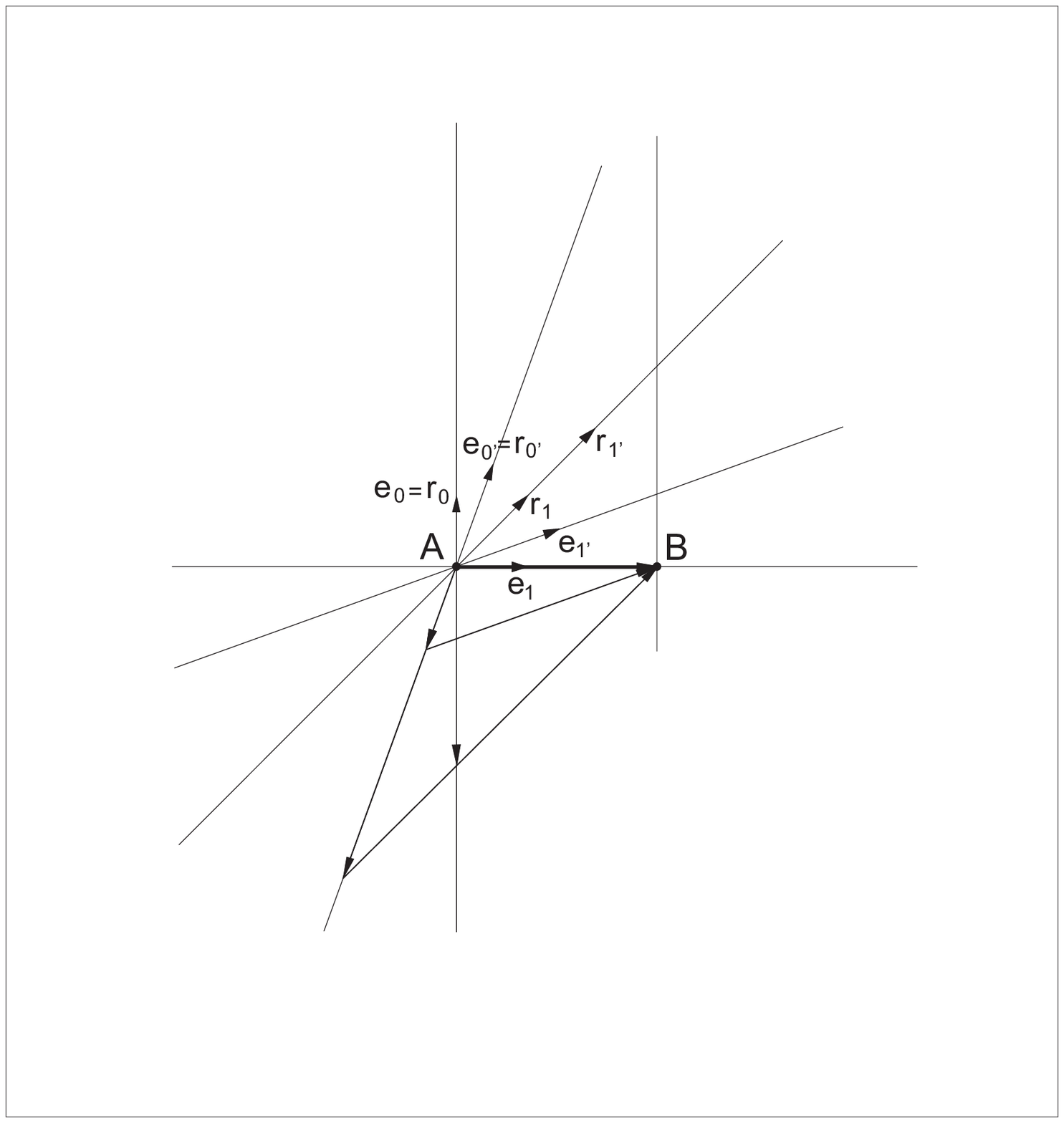}\\[0.5cm]
\end{center}
\noindent Fig.1. The spacetime length for a moving rod. In the ''TT
relativity'' the same quantity for different observers is the geometrical
quantity, the distance 4-vector $l_{AB}^{a}=x_{B}^{a}-x_{A}^{a};$ only \emph{%
one} quantity in 4D spacetime. It is decomposed with respect to $\left\{
e_{\mu }\right\} ,\left\{ e_{\mu ^{\prime }}\right\} $ and $\left\{ r_{\mu
}\right\} ,\left\{ r_{\mu ^{\prime }}\right\} $ bases. The bases $\left\{
e_{\mu }\right\} ,\left\{ e_{\mu ^{\prime }}\right\} $ refer to Einstein's
coordinatization of two relatively moving IFRs $S$ and $S^{\prime },$ and
the bases $\left\{ r_{\mu }\right\} ,\left\{ r_{\mu ^{\prime }}\right\} $
refer to the ''radio'' coordinatization of $S$ and $S^{\prime }.$ $%
l_{AB}^{a} $ corresponds, in the usual ''3+1'' picture, to an object, a rod,
that is at rest in $S$ and situated along the $e_{1}$ basis vector. The
representation of $l_{AB}^{a}$ in the $\left\{ e_{\mu }\right\} $ basis is $%
l_{AB}^{a}=l_{e}^{0}e_{0}+l_{e}^{1}e_{1}=0e_{0}+l_{0}e_{1},$ in the $\left\{
e_{\mu ^{\prime }}\right\} $ basis is $l_{AB}^{a}=-\beta _{e}\gamma
_{e}l_{0}e_{0^{\prime }}+\gamma _{e}l_{0}e_{1^{\prime }},$ in the $\left\{
r_{\mu }\right\} $ basis is $l_{AB}^{a}=-l_{0}r_{0}+l_{0}r_{1},$ and in the $%
\left\{ r_{\mu ^{\prime }}\right\} $ basis is $l_{AB}^{a}=-Kl_{0}r_{0^{%
\prime }}+(1+\beta _{r})(1/K)l_{0}r_{1^{\prime }},$ where $K=(1+2\beta
_{r})^{1/2},$ and $\beta _{r}=\beta _{e}/(1-\beta _{e}).\medskip $
\end{figure}

\noindent \textbf{3.2. The Spacetime Length for a Moving Clock }\medskip

In a similar manner we can choose another particular choice for the distance
4-vector $l_{AB}^{a},$ which will correspond to the well-known ''muon
experiment,'' and which is interpreted in the ''AT relativity'' in terms of
the time ''dilatation''. (This example is also investigated in Ref. 7.)
First we consider this example in the ''TT relativity'' and the situation is
pictured in Fig. 2. The distance 4-vector $l_{AB}^{a}$ will be examined in
two relatively moving IFRs $S$ and $S^{\prime }$ and in the ''e'' and ''r''
coordinatizations, i.e., in the $\left\{ e_{_{\mu }}\right\} ,$ $\left\{
e_{\mu ^{\prime }}\right\} $ and $\left\{ r_{\mu }\right\} ,$ $\left\{
r_{\mu ^{\prime }}\right\} $ bases. The $S$ frame is chosen to be the rest
frame of the muon. Two events are considered; the event $A$ represents the
creation of the muon and the event $B$ represents its decay after the
lifetime $\tau _{0}$ in $S.$ The position 4-vectors of the events $A$ and $B$
in $S$ are taken to be on the world line of a standard clock that is at rest
in the origin of $S.$ The distance 4-vector $l_{AB}^{a}=x_{B}^{a}-x_{A}^{a}$
that connects the events $A$ and $B$ is directed along the $e_{0}$ basis
vector from the event $A$ toward the event $B.$ This geometric quantity can
be written in the coordinate-based geometric language. Thus it can be
decomposed in two bases $\left\{ e_{\mu }\right\} $ and $\left\{ e_{\mu
^{\prime }}\right\} $ as 
\begin{equation}
l_{AB}^{a}=c\tau _{0}e_{0}+0e_{1}=\gamma c\tau _{0}e_{0}^{\prime }-\beta
\gamma c\tau _{0}e_{1}^{\prime }  \label{comu}
\end{equation}
and similarly for the ''r'' coordinatization 
\begin{equation}
l_{AB}^{a}==c\tau _{0}r_{0}+0r_{1}=Kc\tau _{0}r_{0}^{\prime }-\beta
_{r}K^{-1}c\tau _{0}r_{1}^{\prime }  \label{coer}
\end{equation}
\begin{figure}
\begin{center}
\includegraphics[width=8.0cm]{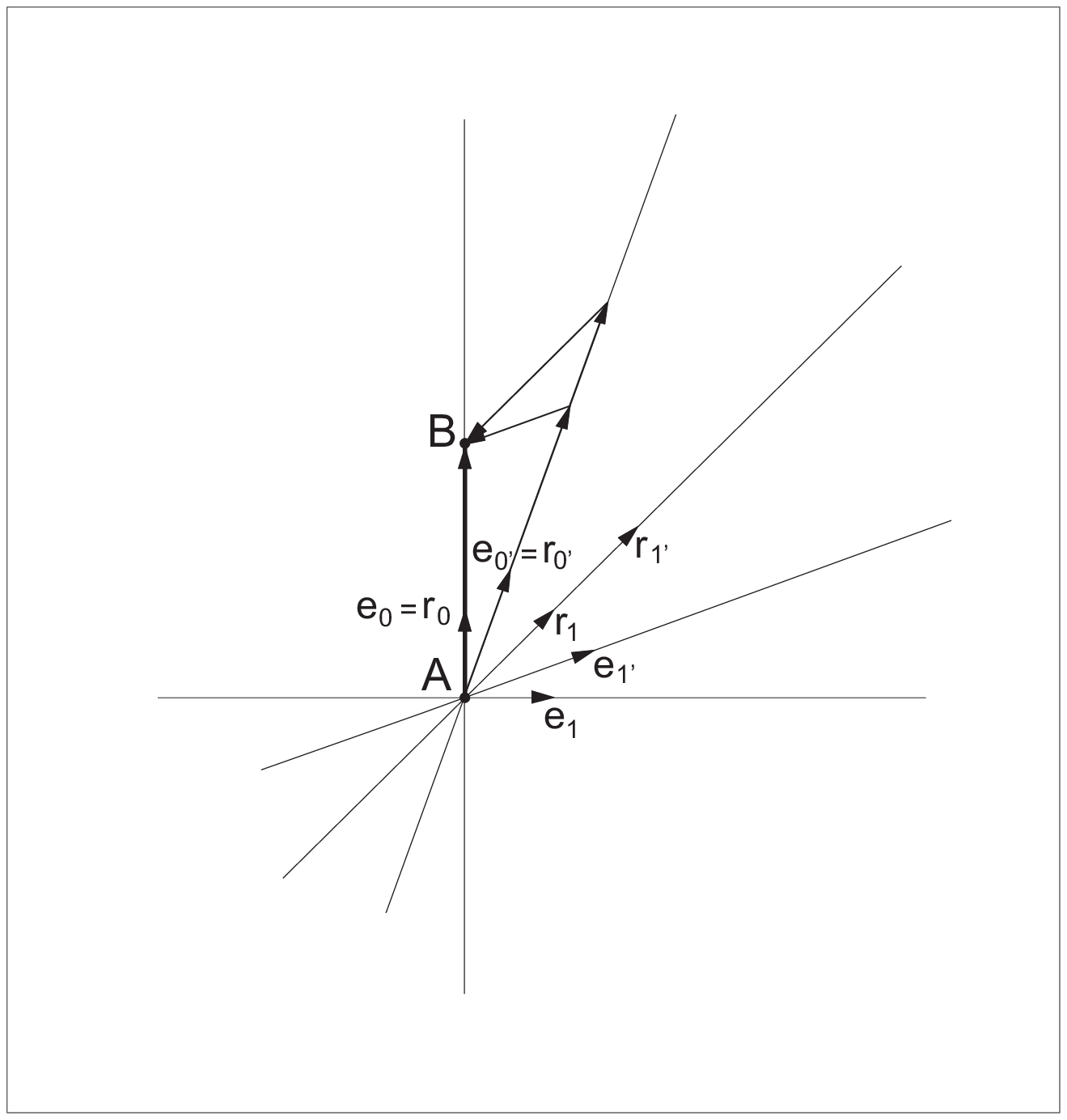}\\[0.5cm]
\end{center}
\noindent Fig.2. The spacetime length for a moving clock. The same
geometrical quantity, the distance 4-vector $l_{AB}^{a}=x_{B}^{a}-x_{A}^{a}$
is decomposed with respect to $\left\{ e_{\mu }\right\} ,\left\{ e_{\mu
^{\prime }}\right\} $ and $\left\{ r_{\mu }\right\} ,\left\{ r_{\mu ^{\prime
}}\right\} $ bases. $l_{AB}^{a}$ connects the events $A$ and $B$ (the event $%
A$ represents the creation of the muon and the event $B$ represents its
decay after the lifetime $\tau _{0}$ in $S$)$.$and it is directed along the $%
e_{0}$ basis vector from the event $A$ toward the event $B.$ The
representation of $l_{AB}^{a}$ in the $\left\{ e_{\mu }\right\} $ basis is $%
l_{AB}^{a}=c\tau _{0}e_{0}+0e_{1},$ in the $\left\{ e_{\mu ^{\prime
}}\right\} $ basis is $l_{AB}^{a}=\gamma c\tau _{0}e_{0^{\prime }}-\beta
\gamma c\tau _{0}e_{1^{\prime }},$ in the $\left\{ r_{\mu }\right\} $ basis
is $l_{AB}^{a}=c\tau _{0}r_{0}+0r_{1},$ and in the $\left\{ r_{\mu ^{\prime
}}\right\} $ basis is $l_{AB}^{a}=Kc\tau _{0}r_{0^{\prime }}-\beta
K^{-1}c\tau _{0}r_{1^{\prime }}.\medskip $
\end{figure}
The equation (\ref{elt5}) also holds for this particular choice of the
distance 4-vector $l_{AB}^{a}.$ We again see that these decompositions,
containing both the basis components and the basis vectors, represent the
same geometric quantity $l_{AB}^{a}.$ $l_{AB}^{a}$ does have only temporal
parts in $S$ (both in the $\left\{ e_{\mu }\right\} $ and $\left\{ r_{\mu
}\right\} $ basis), while in the $\left\{ e_{\mu ^{\prime }}\right\} $ and $%
\left\{ r_{\mu ^{\prime }}\right\} $ basis $l_{AB}^{a}$ contains not only
the temporal part but also the spatial part. Comparing the temporal parts of 
$l_{AB,r}^{\mu }$ and $l_{AB,r}^{\mu ^{\prime }}$ one finds that for some
values of $\beta _{r}$ the temporal part $l_{r}^{0^{\prime }}$ is larger
than $l_{r}^{0}=c\tau _{0}$ and for others it is smaller than $l_{r}^{0}.$
Speaking in the language of the ''AT relativity'' one could say that for
some $\beta _{r}$ there is a time ''dilatation'' while for others $\beta
_{r} $ there is a time ''contraction.'' It is visible from (\ref{comu}) and (%
\ref{coer}) that the comparison of only temporal parts of the
representations of the distance 4-vector is physically meaningless in the
''TT relativity.'' \emph{When only some components of the whole tensor
quantity are taken alone, then, in the ''TT relativity,'' they do not
represent any physical quantity in the 4D spacetime}. The spacetime length $%
l $ is always a well-defined quantity in the ''TT relativity'' and for this
example it is $l=(l_{e}^{\mu }l_{\mu e})^{1/2}=(l_{e}^{\mu ^{\prime }}l_{\mu
^{\prime }e})^{1/2}=(l_{r}^{\mu }l_{\mu r})^{1/2}=(l_{r}^{\mu ^{\prime
}}l_{\mu ^{\prime }r})^{1/2}=(-c^{2}\tau _{0}^{2})^{1/2}$. Since in $S$ the
spatial parts $l_{e,r}^{1}$ of $l_{e,r}^{\mu }$ are zero the spacetime
length $l$ in $S$ is a measure of the temporal distance, as in the
prerelativistic physics; one defines that $c^{2}\tau _{0}^{2}=-l_{e}^{\mu
}l_{\mu e}=-l_{r}^{\mu }l_{\mu r}.\bigskip \medskip $

\noindent \textbf{4. THE\ AT\ OF\ SPATIAL\ AND\ TEMPORAL\ DISTANCES}\
\bigskip

In this section we consider the same two examples as above but now from the
point of view of the conventional, i.e., Einstein's$^{(1)}$ interpretations
of \emph{the} \emph{spatial length} of the moving rod and \emph{the temporal
distance} for the moving clock.\bigskip \medskip

\noindent \textbf{4.1. The AT of the Spatial Distance} \medskip

The AT of the spatial distance is already considered in the usual ''AT
relativity'' approach in Refs. 3 and 7, and therefore, here, we only quote
the main results and the definitions, and also illustrate the whole
consideration by Fig. 3. The same example, a rod at rest in $S,$ is pictured
in Fig. 1 when treated in the ''TT relativity,'' and in Fig. 3 when treated
in the ''AT relativity.'' It is mentioned in Ref. 3 that the synchronous
definition of \emph{the spatial length}, introduced by Einstein$^{(1)}$
defines length as \emph{the spatial distance} between two spatial points on
the (moving) object measured by simultaneity in the rest frame of the
observer. In the ''AT relativity'' the concept of sameness of a physical
quantity differs from that one in the ''TT relativity.'' Indeed, in the
usual ''AT relativity'' one takes only some basis components of $l_{AB}^{a}$%
\ (that is, of the CBGQs $l_{e}^{\mu }e_{\mu }$\ and $l_{e}^{\mu ^{\prime
}}e_{\mu ^{\prime }}$) in $S$\ and $S^{\prime },$\ then performs some
additional manipulations with them, and considers that the constructed
quantities represent the same physical quantity for observers in $S$\ and $%
S^{\prime }$. Thus for the Einstein's definition of \emph{the spatial length}
one considers only \emph{the component} $l_{e}^{1}=l_{0}$ of $l_{e}^{\mu
}e_{\mu }$ (when $l_{e}^{0}$ is taken $=0,$ i.e., the spatial ends of the
rod at rest in $S$ are taken simultaneously at $t=0$). Further one compares
it with the quantity which is obtained in the following way; first one
performs the Lorentz transformation $L^{\mu }{}_{\nu ^{\prime },e}$ of the
basis components $l_{e}^{\nu ^{\prime }}$ (but not of the basis itself) from 
$S^{\prime }$ to $S,$ which yields 
\begin{eqnarray}
l_{e}^{0} &=&\gamma _{e}l_{e}^{0^{\prime }}+\gamma _{e}\beta
_{e}l_{e}^{1^{\prime }}  \nonumber \\
l_{e}^{1} &=&\gamma _{e}l_{e}^{1^{\prime }}+\gamma _{e}\beta
_{e}l_{e}^{0^{\prime }}  \label{elcon}
\end{eqnarray}
Then \emph{one retains only the transformation of the spatial component }$%
l_{e}^{1}$\emph{\ }(the second equation in (\ref{elcon})) \emph{neglecting
completely the transformation of the temporal part} $l_{e}^{0}$ (the first
equation in (\ref{elcon})). From the ''TT relativity'' viewpoint this step
of derivation is unjustified and, in fact, incorrect. Furthermore in the
transformation for $l_{e}^{1}$ one takes that the temporal part in $%
S^{\prime }$ $l_{e}^{0^{\prime }}=0,$ ( i.e., the spatial ends of the rod
moving in $S^{\prime }$ are taken simultaneously at some \emph{arbitrary }$%
t^{\prime }=b$). Again an incorrect step from the ''TT relativity''
viewpoint. The quantity obtained in such a way will be denoted as $%
L_{e}^{1^{\prime }}$ (it is not equal to $l_{e}^{1^{\prime }}$ appearing in
the transformation equations (\ref{elcon})). This quantity $L_{e}^{1^{\prime
}}$ defines in the ''AT relativity'' \emph{the synchronously determined
spatial length }of the moving rod in $S^{\prime }$ (in Fig. 3 $%
L_{e}^{1^{\prime }}=x_{De}^{1^{\prime }}-x_{Ce}^{1^{\prime }}$). The
mentioned procedure gives $l_{e}^{1}=\gamma _{e}L_{e}^{1^{\prime }},$ that
is, the famous formula for \emph{the Lorentz ''contraction,''} 
\begin{equation}
L_{e}^{1^{\prime }}=l_{e}^{1}/\gamma _{e}=l_{0}/\gamma _{e}  \label{apcon}
\end{equation}
This quantity, $L_{e}^{1^{\prime }}=l_{0}/\gamma _{e},$ is the usual Lorentz
contracted \emph{spatial length}$,$ and \emph{the quantities }$l_{0}$\emph{\
and }$L_{e}^{1^{\prime }}=l_{0}/\gamma _{e}$\emph{\ are considered in the
''AT relativity'' to be the same quantity for observers in }$S$\emph{\ and }$%
S^{\prime }$\emph{.} The comparison with the relation (\ref{trucon}) (and
Fig. 3) clearly shows that the constructed quantities $l_{0}$\emph{\ and }$%
L_{e}^{1^{\prime }}=l_{0}/\gamma _{e}$\emph{\ are two different and
independent quantities in 4D spacetime.} Namely, these quantities are
obtained by the same measurements in $S$ and $S^{\prime };$ the spatial ends
of the rod are measured simultaneously at some $t_{e}=a$ in $S$ (in Fig. 3 $%
t_{e}=t_{Be}=t_{Ae}=0$) and also at some $t_{e}^{\prime }=b$ in $S^{\prime }$
(in Fig. 3 $t_{e}^{\prime }=t_{De}^{\prime }=t_{Ce}^{\prime }=b$). $a$ in $S$
and $b$ in $S^{\prime }$ are not related by the Lorentz transformation $%
L^{\mu }{}_{\nu ,e}$ or any other coordinate transformation. The same
happens in the ''r'' coordinatization, where the analogous procedure yields
the relation between $L_{r}^{1^{\prime }}$ and $l_{r}^{1}=l_{0}$ as the
Lorentz ''contraction'' of the moving rod in the ''r'' coordinatization 
\begin{equation}
L_{r}^{1^{\prime }}=l_{0}/K=(1+2\beta _{r})^{-1/2}l_{0}  \label{aper}
\end{equation}
\begin{figure}
\begin{center}
\includegraphics[width=8.0cm]{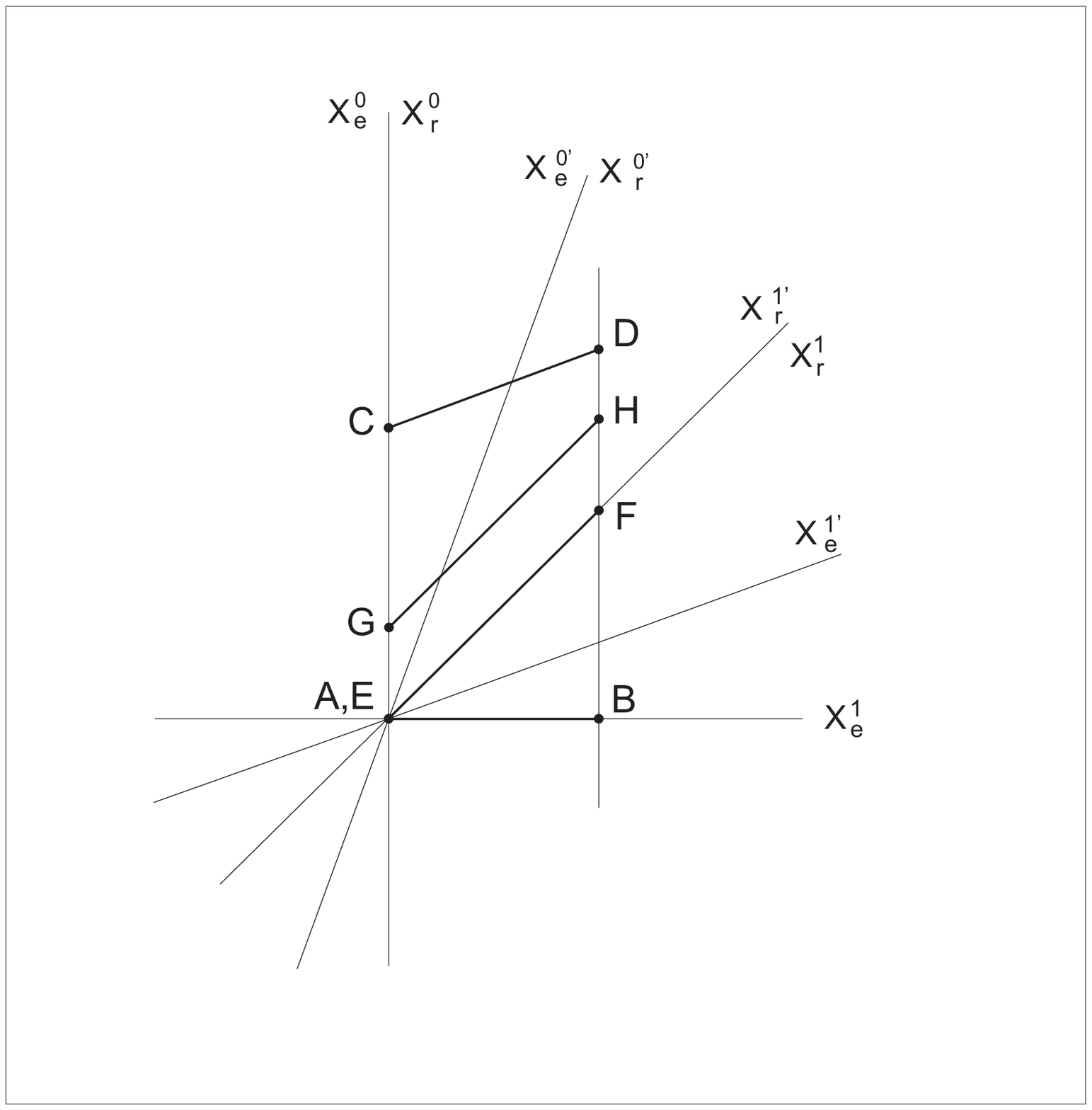}\\[0.5cm]
\end{center}
\noindent Fig.3. The AT of the spatial length - the Lorentz ''contraction''
of the moving rod. The spatial distance $%
l_{ABe}^{1}=x_{Be}^{1}-x_{Ae}^{1}=l_{e}^{1}=l_{0}$ defines in the ''AT
relativity,'' and in the ''e'' coordinatization, the spatial length of the
rod at rest in $S,$ while $L_{e}^{1^{\prime }}=x_{De}^{1^{\prime
}}-x_{Ce}^{1^{\prime }}$ is considered in the ''AT relativity,'' and in the
''e'' basis, to define the spatial length of the moving rod in $S^{\prime }.$
$L_{e}^{1^{\prime }}$ and $l_{e}^{1}=l_{0}$ are connected by the formulae
for the Lorentz contraction of the moving rod $L_{e}^{1^{\prime
}}=l_{0}/\gamma _{e},$ with $t_{Ce}^{\prime }=t_{De}^{\prime }=t_{e}^{\prime
}=b$ and $t_{Be}=t_{Ae}=t_{e}=a.$ $a$ in $S$ and $b$ in $S^{\prime }$ are
not related by the Lorentz transformation or any other coordinate
transformation. Likewise in the ''r'' coordinatization, the spatial distance 
$l_{EFr}^{1}=x_{Fr}^{1}-x_{Er}^{1}$\ defines in the ''AT relativity'' the
spatial length of the rod at rest in $S,$\ while $L_{r}^{1^{\prime
}}=x_{Hr}^{1^{\prime }}-x_{Gr}^{1^{\prime }}$\ defines the spatial length of
the moving rod in $S^{\prime }$. $L_{r}^{1^{\prime }}$ and $%
l_{r}^{1}=l_{EFr}^{1}=l_{0}$ are connected by the formulae for the Lorentz
''contraction'' of the moving rod in the ''r'' coordinatization $%
L_{r}^{1^{\prime }}=l_{0}/K$ with $x_{Hr}^{0^{\prime }}=x_{Gr}^{0^{\prime }}$
and $x_{Fr}^{0}=x_{Er}^{0}.$ In the ''r'' coordinatization there is a length
dilatation $\infty \succ L_{r}^{1^{\prime }}\succ l_{0}$ for $-1/2\prec
\beta _{r}\prec 0$ and the standard ''length contraction'' $l_{0}\succ
L_{r}^{1^{\prime }}\succ 0$ for positive $\beta _{r},$ which clearly shows
that the ''Lorentz contraction'' is not physically correctly defined
transformation. In the ''AT relativity'' all four spatial lengths $%
l_{e}^{1}, $ $L_{e}^{1^{\prime }},$ $l_{r}^{1},$ $L_{r}^{1^{\prime }}$ are
considered as the same quantity for different observers, but, in fact, they
are four different quantities in 4D spacetime, and they are not connected by
the Lorentz transformation.\medskip
\end{figure}
Let us explain this result in more detail. The spatial ends of the
considered rod, which is at rest in $S,$ must be taken simultaneously in the
''r'' coordinatization as well. Thus they must lie on the light line, i.e.,
on the $x_{r}^{1}$ axis (that is along the spatial base vector $r_{1}$). The
simultaneous events $E$ and $F$ (whose spatial parts correspond to the
spatial ends of the rod) are the intersections of the $x_{r}^{1}$ axis and
the world lines of the spatial ends of the rod. The events $E$ and $F$ are
not the same events as the events $A$ and $B,$ considered in the ''e''
coordinatization for the same rod at rest in $S,$ since the simultaneity of
the events is defined in different ways, see Fig.3. Therefore, in 4D
spacetime the spatial length in the $\left\{ r_{\mu }\right\} $ basis $%
l_{r}^{1}=l_{0}$ (with $x_{Fr}^{0}=x_{Er}^{0}$) is not the same 4D quantity
as the spatial length in the $\left\{ e_{\mu }\right\} $ basis $%
l_{e}^{1}=l_{0}$ (with $x_{Be}^{0}=x_{Ae}^{0}).$ Applying the same procedure
as above (and in Ref. 3) one finds that in the ''r'' coordinatization, the
spatial distance $l_{r}^{1}=x_{Fr}^{1}-x_{Er}^{1}=l_{0}$\ (with $%
x_{Fr}^{0}=x_{Er}^{0})$ defines in the ''AT relativity'' the spatial length
of the rod at rest in $S,$\ while $L_{r}^{1^{\prime }}=x_{Hr}^{\prime
1}-x_{Gr}^{\prime 1}$\ (with $x_{Hr}^{0^{\prime }}=x_{Gr}^{0^{\prime }}$)
defines the spatial length of the moving rod in $S^{\prime },$ see Fig. 3.
We see from (\ref{aper}) that there is a length dilatation $\infty \succ
L_{r}^{1^{\prime }}\succ l_{0}$ for $-1/2\prec \beta _{r}\prec 0$ and the
standard length ''contraction'' $l_{0}\succ L_{r}^{1^{\prime }}\succ 0$ for
positive $\beta _{r},$ which clearly shows that the Lorentz ''contraction''
is not a physically correctly defined transformation. Thus, \emph{in the
''TT relativity'' the same quantity for different observers is the tensor
quantity, the 4-vector }$l_{AB}^{a}=l_{e}^{\mu }e_{\mu }=l_{e}^{\mu ^{\prime
}}e_{\mu ^{\prime }}=l_{r}^{\mu }r_{\mu }=l_{r}^{\mu ^{\prime }}r_{\mu
^{\prime }};$\emph{\ only one quantity in 4D spacetime.} However \emph{in
the ''AT relativity'' different quantities in 4D spacetime, the spatial
distances }$l_{e}^{1},$\emph{\ }$L_{e}^{1^{\prime }},$\emph{\ }$l_{r}^{1},$%
\emph{\ }$L_{r}^{1^{\prime }},$\emph{\ are considered as the same quantity
for different observers. }It is also shown in Ref. 3 that the Lorentz
''contraction'' as the coordinate transformation changes the infinitesimal
interval $ds,$ which defines the geometry of spacetime. \emph{Thus the
Lorentz contraction is the transformation connecting different quantities in 
}$S$\emph{\ and }$S^{\prime }$\emph{\ and changing }$ds,$\emph{\ which
implies that it is an AT. \bigskip \medskip }

\noindent \textbf{4.2. The AT of the Temporal Distance} \medskip

The same example of the ''muon decay'' will be now considered in the ''AT
relativity'' (see also Ref. 7). In the ''e'' coordinatization the events $A$
and $B$ are again on the world line of a muon that is at rest in $S$ as
depicted in Fig. 4. We shall see once again that the concept of sameness of
a physical quantity is quite different in the ''AT relativity.'' There one
takes only some components of $l_{AB}^{a}$\ (that is, of the CBGQs $%
l_{e}^{\mu }e_{\mu }$\ and $l_{e}^{\mu ^{\prime }}e_{\mu ^{\prime }}$) in $S$%
\ and $S^{\prime },$\ then performs some additional manipulations with them,
and considers that the constructed quantities represent the same physical
quantity for observers in two relatively moving IFRs $S$\ and $S^{\prime }$.
Thus for this example one compares \emph{the basis component} $%
l_{e}^{0}=c\tau _{0}$ of $l_{e}^{\mu }e_{\mu }$ with the quantity, which is
obtained from \emph{the basis component} $l_{e}^{0^{\prime }}$ in the
following manner; first one performs the Lorentz transformation of the basis
components $l_{e}^{\mu }$ (but not of the basis itself) from the muon rest
frame $S$ to the frame $S^{\prime }$ in which the muon is moving. This
procedure yields 
\begin{eqnarray}
l_{e}^{0^{\prime }} &=&\gamma _{e}l_{e}^{0}-\gamma _{e}\beta _{e}l_{e}^{1} 
\nonumber \\
l_{e}^{1^{\prime }} &=&\gamma _{e}l_{e}^{1}-\gamma _{e}\beta _{e}l_{e}^{0}
\label{eltime}
\end{eqnarray}
Similarly as in the Lorentz contraction \emph{one now forgets the
transformation of the spatial part} $l_{e}^{1^{\prime }}$ (the second
equation in (\ref{eltime})) \emph{and considers only the transformation of
the temporal part} $l_{e}^{0^{\prime }}$ (the first equation in (\ref{eltime}%
)). This is, of course, an incorrect step from the ''TT relativity''
viewpoint. Then taking that $l_{e}^{1}=0$ (i.e., that $x_{Be}^{1}=x_{Ae}^{1}$%
) in the equation for $l_{e}^{0^{\prime }}$ (the first equation in (\ref
{eltime})) one finds the new quantity which will be denoted as $%
L_{e}^{0^{\prime }}$ (it is not the same as $l_{e}^{0^{\prime }}$ appearing
in the transformation equations (\ref{eltime})). The temporal distance $%
l_{e}^{0}$ defines in the ''AT relativity,'' and in the ''e''
coordinatization, the muon lifetime at rest, while $L_{e}^{0^{\prime }}$ is
considered in the ''AT relativity,'' and in the ''e'' coordinatization, to
define the lifetime of the moving muon in $S^{\prime }.$ The relation
connecting $L_{e}^{0^{\prime }}$ with $l_{e}^{0},$ which is obtained by the
above procedure, is then the well-known relation for \emph{the ''time
dilatation,''} 
\begin{equation}
L_{e}^{0^{\prime }}/c=t_{e}^{\prime }=\gamma _{e}l_{e}^{0}/c=\tau
_{0}(1-\beta _{e}^{2})^{-1/2}  \label{tidil}
\end{equation}
Analogously we find in the ''r'' coordinatization that 
\begin{equation}
L_{r}^{0^{\prime }}=Kl_{r}^{0}=(1+2\beta _{r})^{1/2}c\tau _{0}  \label{tider}
\end{equation}
This relation shows that the new quantity $L_{r}^{0^{\prime }},$ which
defines in the ''AT relativity'' the temporal separation in $S^{\prime },$
where the clock is moving, is smaller - time ''contraction'' - than the
temporal separation $l_{r}^{0}=c\tau _{0}$ in $S,$ where the clock is at
rest, for $-1/2\prec \beta _{r}\prec 0,$ and it is larger - time
''dilatation'' - for $0\prec \beta _{r}\prec \infty $.

>From this consideration we conclude that \emph{in the ''TT relativity'' the
same quantity for different observers is the tensor quantity, the 4-vector }$%
l_{AB}^{a}=l_{e}^{\mu }e_{\mu }=l_{e}^{\mu ^{\prime }}e_{\mu ^{\prime
}}=l_{r}^{\mu }r_{\mu }=l_{r}^{\mu ^{\prime }}r_{\mu ^{\prime }};$\emph{\
only one quantity in 4D spacetime.} However \emph{in the ''AT relativity''
different quantities in 4D spacetime, the temporal distances }$l_{e}^{0},$%
\emph{\ }$L_{e}^{0^{\prime }},$\emph{\ }$l_{r}^{0},$\emph{\ }$%
L_{r}^{0^{\prime }},$\emph{\ are considered as the same quantity for
different observers. This shows that the time ''dilatation'' is the
transformation connecting different quantities in }$S$\emph{\ and }$%
S^{\prime }$\emph{\ and therefore it is an AT. }

The considerations from Secs. 4.1 and 4.2 reveal that \emph{both the Lorentz
''contraction'' and the time ''dilatation'' are the transformations
connecting different quantities (in 4D spacetime) in different IFRs, and
also they both change the infinitesimal spacetime distance }$ds,$\emph{\
i.e., the pseudo-Euclidean geometry of the 4D spacetime }(this is explicitly
shown in Ref. 3 for the Lorentz ''contraction,'' and it can be easily shown
for the time ''dilatation''). Thence \emph{\ both transformations the
Lorentz ''contraction'' and the time ''dilatation'' belong to - the AT.}%
\textit{\ }

We can compare the obtained results for the determination of the spacetime
length in the ''TT relativity'' and the determination of the spatial and
temporal distances in the ''AT\ relativity'' with the existing experiments.
This comparison is presented in Ref. 9. It is shown there that the ''TT
relativity'' results agree with all experiments that are complete from the
''TT relativity'' viewpoint, i.e., in which all parts of the considered
tensor quantity are measured in the experiment. However the ''AT
relativity'' results agree only with some of the examined
experiments.\bigskip \medskip
\begin{figure}
\begin{center}
\includegraphics[width=8.0cm]{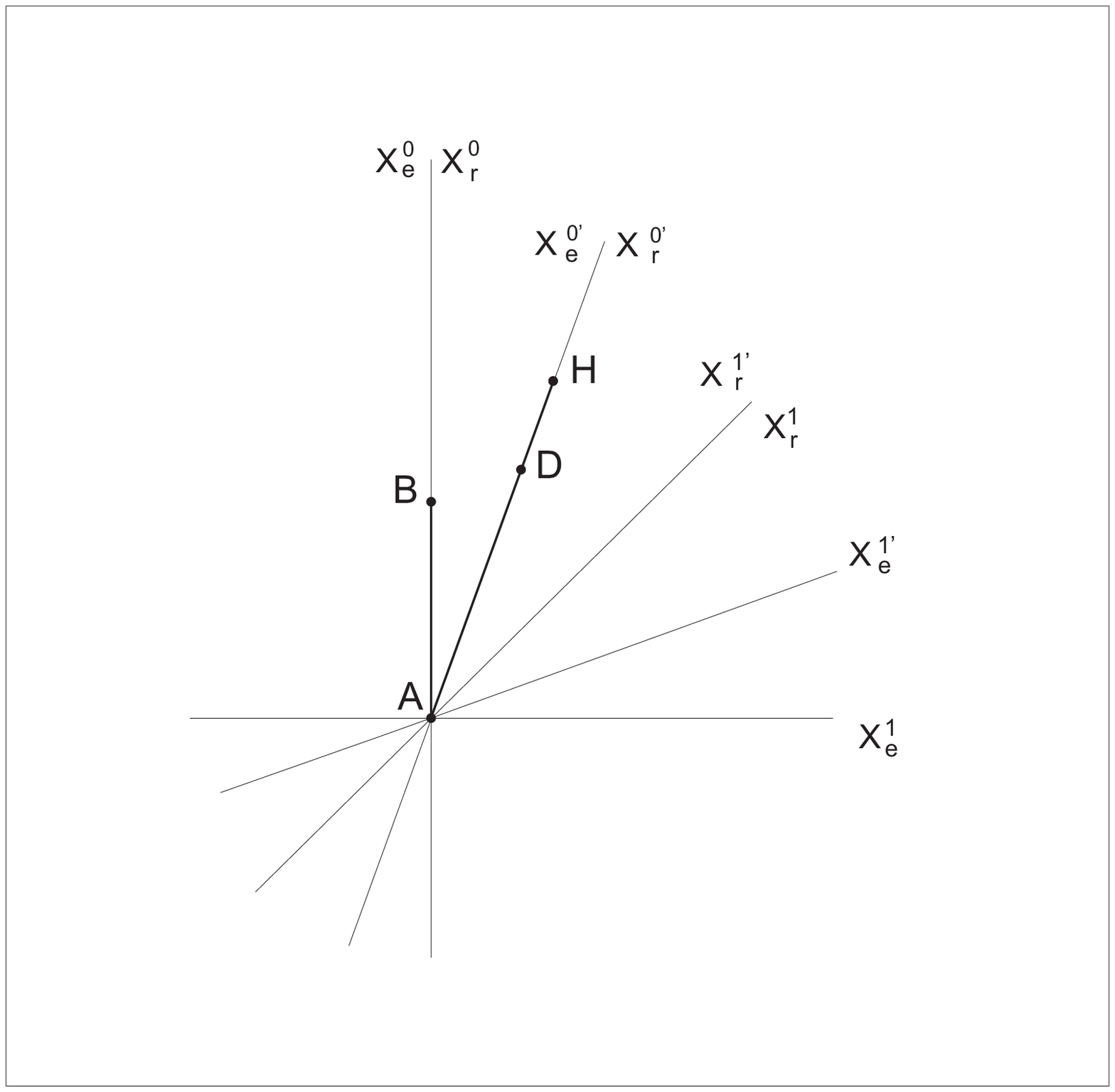}\\[0.5cm]
\end{center}
\noindent Fig.4. The AT of the temporal distance - the ''dilatation'' of
time for the moving clock. The temporal distance $l_{ABe}^{0}=l_{e}^{0}$
defines in the ''AT relativity,'' and in the ''e'' coordinatization, the
muon lifetime at rest, while $L_{e}^{0^{\prime }}$ is considered in the ''AT
relativity,'' and in the ''e'' coordinatization, to define the lifetime of
the moving muon in $S^{\prime }.$ The quantities $L_{e}^{0^{\prime }}$ and $%
l_{e}^{0}$ are connected by the relation for the time dilatation, $%
L_{e}^{0^{\prime }}/c=t_{e}^{\prime }=\gamma _{e}l_{e}^{0}/c=\tau
_{0}(1-\beta _{e}^{2})^{-1/2},$ with $x_{Be}^{1}=x_{Ae}^{1}.$ Likewise, the
temporal distance $l_{ABr}^{0}=l_{r}^{0}$ defines in the ''AT relativity,''
and in the ''r'' coordinatization, the muon lifetime at rest, while $%
L_{r}^{0^{\prime }}$ is considered in the ''AT relativity,'' and in the
''r'' coordinatization, to define the lifetime of the moving muon in $%
S^{\prime }.$ $L_{r}^{0^{\prime }}$ and $l_{r}^{0}$ are connected by the
relation for the time ''dilatation'' in the ''r'' coordinatization $%
L_{r}^{0^{\prime }}=Kl_{r}^{0}=(1+2\beta _{r})^{1/2}c\tau _{0}.$ The
temporal separation $L_{r}^{0^{\prime }}$ in $S^{\prime },$ where the clock
is moving, is smaller - ''time contraction'' - than the temporal separation $%
l_{r}^{0}=c\tau _{0}$ in $S,$ where the clock is at rest, for $-1/2\prec
\beta _{r}\prec 0,$ and it is larger - ''time dilatation'' - for $0\prec
\beta _{r}\prec \infty $. The ''AT relativity'' considers the temporal
distances $l_{e}^{0},$\emph{\ }$L_{e}^{0^{\prime }},$\emph{\ }$l_{r}^{0},$%
\emph{\ }$L_{r}^{0^{\prime }}$ as the same quantity for different observers.
However these temporal distances are really different quantities in 4D
spacetime, and they are not connected by the Lorentz transformations.
\end{figure}

\noindent \textbf{5. THE\ AT\ OF\ THE\ 3-VECTORS }$\mathbf{E}$\textbf{\ AND }%
$\mathbf{B}$ \bigskip

It is generally believed that the covariant formulation of the
electrodynamics with $F^{\alpha \beta }$ (the component form) and the usual
formulation with $\mathbf{E}$ and $\mathbf{B}$ are equivalent, and therefore
that the usual transformations of $\mathbf{E}$ and $\mathbf{B}$ are actually
the TT. However it is revealed in Refs. 2, 3 and 17 that the usual
transformations of the 3-vectors $\mathbf{E}$ and $\mathbf{B}$ are also the
AT. Here we shall present another proofs that the conventional
transformations of $\mathbf{E}$ and $\mathbf{B}$ are the AT. Furthermore it
will be shown in this section that the Maxwell equations in the 3-vector
form are not equivalent to the Maxwell equations formulated as the true
tensor equations or as the CBGEs.\bigskip \medskip

\noindent \textbf{5.1. The Usual Derivation of the Transformations of }$%
\mathbf{E}$\textbf{\ and }$\mathbf{B}$\textbf{\ by Using} $F^{\alpha \beta }$
\medskip

The derivation of the usual transformations of the 3-vectors $\mathbf{E}$
and $\mathbf{B}$ which is based on the AT of the spatial distance, i.e., the
Lorentz ''contraction'' (see, for example, Ref. 18), will be presented at
the end of Sec. 5.3. Here we start with a simple derivation of the
transformation relations for $\mathbf{E}$ and $\mathbf{B}$ by using $%
F^{\alpha \beta }$ (for similar derivation see, e.g., Ref. 10 Sec. 11.10,
Ref. 13 Sec. 3.3.). It has to be noted that \emph{such conventional
derivation is made using the component form of tensor quantities and
equations, and the components are determined in the specific
coordinatization, the Einstein coordinatization.} (Because of that we shall
omit the subscript $^{\prime }e^{\prime }$ for the ''e'' coordinatization in
the rest part of this paper, but we still denote the quantities in the ''r''
coordinatization by the subscript $^{\prime }r^{\prime }.$) First one
identifies, in some IFR $S$, the components $E_{i}$ and $B_{i}$ of the
3-vectors $\mathbf{E}$ and $\mathbf{B}$ with \emph{some of the basis
components of} $F^{ab}$ as 
\begin{equation}
E_{i}=F^{0i},\quad B_{i}=(1/c)^{*}F^{0i}  \label{eibi}
\end{equation}
in order to get in that IFR the usual Maxwell equations, 
\begin{eqnarray}
\nabla \mathbf{E}(\mathbf{r},t) &=&\rho (\mathbf{r},t)/\varepsilon
_{0},\quad \nabla \times \mathbf{E}(\mathbf{r},t)=-\partial \mathbf{B}(%
\mathbf{r},t)/\partial t  \nonumber \\
\nabla \mathbf{B}(\mathbf{r},t) &=&0,\quad \nabla \times \mathbf{B}(\mathbf{r%
},t)=\mu _{0}\mathbf{j}(\mathbf{r},t)+\mu _{0}\varepsilon _{0}\partial 
\mathbf{E}(\mathbf{r},t)/\partial t  \label{max}
\end{eqnarray}
from the covariant Maxwell equations with $F^{\alpha \beta }$ and its dual $%
^{*}F^{\alpha \beta }$ 
\begin{equation}
\partial _{\alpha }F^{a\beta }=-j^{\beta }/\varepsilon _{0}c,\quad \partial
_{\alpha }\ ^{*}F^{\alpha \beta }=0  \label{maxco}
\end{equation}
where $^{*}F^{\alpha \beta }=-(1/2)\varepsilon ^{\alpha \beta \gamma \delta
}F_{\gamma \delta }$ and $\varepsilon ^{\alpha \beta \gamma \delta }$ is the
totally skew-symmetric Levi-Civita pseudotensor. The equations (\ref{maxco})
are written in the component form and in the ''e'' coordinatization.

After transforming by the Lorentz transformation $L^{\mu ^{\prime }}{}_{\nu
,e}$ (\ref{lorus}) the covariant Maxwell equations for basis components (\ref
{maxco}) to the $S^{\prime }$ frame one finds 
\begin{equation}
\partial _{\alpha ^{\prime }}F^{a^{\prime }\beta ^{\prime }}=-j^{\beta
^{\prime }}/\varepsilon _{0}c,\quad \partial _{\alpha ^{\prime }}\
^{*}F^{\alpha ^{\prime }\beta ^{\prime }}=0  \label{maxc1}
\end{equation}
We note that we could directly write the equations (\ref{maxc1}) by using
the following rule (see Ref. 12, Sec. 6.1): ''If an equation is formed using
components of tensors combined only by \emph{the permissible tensor
operations} (my emphasis), and if the equation is true in one basis, then it
is true in any other.'' In this case, if the equation (\ref{maxco}) is valid
in the $\left\{ e_{\mu }\right\} $ basis then it will be valid in the $%
\left\{ e_{\mu ^{\prime }}\right\} $ basis, (\ref{maxc1}), as well, since (%
\ref{maxco}) is formed combining components of tensors by the permissible
tensor operations. Then, in $S^{\prime },$ one again identifies the
components $E_{i^{\prime }}$ and $B_{i^{\prime }}$ of the 3-vectors $\mathbf{%
E}^{\prime }$ and $\mathbf{B}^{\prime }$ with \emph{some of the basis
components of} $F^{ab}$ \emph{in the same way as in} $S$, i.e., 
\begin{equation}
E_{i^{\prime }}=F^{0^{\prime }i^{\prime }},\quad B_{i^{\prime
}}=(1/c)^{*}F^{0^{\prime }i^{\prime }}  \label{eib1}
\end{equation}
in order to obtain in that frame the usual Maxwell equations (in the
3-vector form) from the transformed covariant Maxwell equations for basis
components. This procedure then gives the connection between the quantities $%
E_{i^{\prime }},$ $B_{i^{\prime }}$ in $S^{\prime }$ and $E_{i},$ $B_{i}$ in 
$S$ as 
\begin{eqnarray}
E_{i^{\prime }} &=&\Gamma (E_{i}+c\varepsilon _{ijk}\beta
_{j}B_{k})-((\Gamma -1)/\beta ^{2})\beta _{i}(\beta _{k}E_{k})  \nonumber \\
B_{i^{\prime }} &=&\Gamma (B_{i}-(1/c)\varepsilon _{ijk}\beta
_{j}E_{k})-((\Gamma -1)/\beta ^{2})\beta _{i}(\beta _{k}B_{k})
\label{eitran}
\end{eqnarray}
where $\beta =V/c$ and $\Gamma =(1-\beta ^{2})^{-1/2}.$ When (\ref{eitran})
is written in terms of the chosen basis components of $F^{ab},$ then, e.g.,
the first equation in (\ref{eitran}) becomes 
\begin{equation}
F^{0^{\prime }i^{\prime }}=\Gamma (F^{0i}+(1/2)\varepsilon _{ijk}\varepsilon
_{klm}\beta _{j}F^{lm})-((\Gamma -1)/\beta ^{2})\beta _{i}(\beta _{k}F^{0k})
\label{efeic}
\end{equation}
The components of the 3-vector fields $\mathbf{E}$ and $\mathbf{B}$, and of
the 3-velocity $\mathbf{V}$ are written with lowered (generic) subscripts,
since they are not the spatial components of the 4-vectors. This refers to
the third-rank antisymmetric $\varepsilon $ tensor too. The super- and
subscripts are used only on the components of the 4-vectors or tensors.

Now, from the quantities $E_{i},$ $E_{i^{\prime }}$, $B_{i}$ and $%
B_{i^{\prime }}$ \emph{and the basis 3-vectors in the 3D space }$\mathbf{e}%
_{i}$ and $\mathbf{e}_{i^{\prime }}$ one constructs the electric and
magnetic fields 3-vectors as $\mathbf{E=}E_{i}\mathbf{e}_{i},$ $\mathbf{B}%
=B_{i}\mathbf{e}_{i}$ in $S$ and $\mathbf{E}^{\prime }\mathbf{=}E_{i^{\prime
}}\mathbf{e}_{i^{\prime }},$ $\mathbf{B}^{\prime }=B_{i^{\prime }}\mathbf{e}%
_{i^{\prime }}$ in $S^{\prime }.$ Then the equations for components (\ref
{eitran}) (actually the equation (\ref{efeic})) are written as the relations
between the 3-vectors 
\begin{eqnarray}
\mathbf{E}^{\prime } &=&\Gamma (\mathbf{E}+c\mathbf{\beta }\times \mathbf{B}%
)-((\Gamma -1)/\beta ^{2})\mathbf{\beta }(\mathbf{\beta E)}  \nonumber \\
\mathbf{B}^{\prime } &=&\Gamma (\mathbf{B}-c\mathbf{\beta }\times \mathbf{E}%
)-((\Gamma -1)/\beta ^{2})\mathbf{\beta }(\mathbf{\beta B)}  \label{eitr2}
\end{eqnarray}
In Refs. 10 and 13 a little different procedure is chosen to find the
relation (\ref{eibi}). In fact, they first made the identification of the
basis components of $F^{ab}$ with the components of the 3-vectors $\mathbf{E}
$ and $\mathbf{B}$, and then the relation (\ref{eitran}) was obtained using
the relations (\ref{eib1}) and (\ref{efeic}).

An important remark is in place already here. The identification of the
components of the 3-vectors $\mathbf{E}$ and $\mathbf{B}$ with \emph{some}
of the basis components of $F^{ab},$ the equation (\ref{eibi}), \emph{is not
the permissible tensor operation}. This means that if the usual Maxwell
equations with $\mathbf{E}$ and $\mathbf{B}$ (\ref{max}), which are obtained
by such an operation, are true in one basis (a specific Lorentz frame $S$
and the ''e'' coordinatization), then they do not need to be true in another
basis, i.e., in the $S^{\prime }$ frame. There is no rule inside the tensor
calculus which can guarantee that the validity of the usual Maxwell
equations with $\mathbf{E}^{\prime }$ and $\mathbf{B}^{\prime }$ results
from the validity of the corresponding Maxwell equations (\ref{max}) with $%
\mathbf{E}$ and $\mathbf{B.}$

Einstein's fundamental work$^{(19)}$ is the earliest reference on generally
covariant electrodynamics and on the identification of some components of $%
F^{ab}$ (actually $F^{\alpha \beta }$) with the components of $\mathbf{E}$
and $\mathbf{B.}$ He introduces an electromagnetic potential 4-vector (in
component form) and from this constructs $F^{a\beta },$ the component form
of the $F^{ab}$ tensor. Then he writes the equations (\ref{maxco}) and shows
that these equations correspond to the usual Maxwell equations with $\mathbf{%
E}$ and $\mathbf{B}$ if he makes the identification given in the equations (%
\ref{eibi}). It has to be mentioned that Einstein worked with \emph{the
equations for basis components in the ''e'' coordinatization }and thus not
with the tensor equations, or with the CBGEs (see Ref. 16 for the comparison
of Einstein's view of spacetime and the modern view).\bigskip \medskip

\noindent \textbf{5.2. The Discussion of the Usual Derivation of the
Transformations of }$\mathbf{E}$\textbf{\ and }$\mathbf{B}$\textbf{\ from
the ''TT Relativity'' Viewpoint\medskip }

Let us discuss such conventional derivations of the relations (\ref{eitran})
and (\ref{eitr2}) from the point of view of the ''TT relativity.'' First of
all $F^{\alpha \beta }$ and $F^{\alpha ^{\prime }\beta ^{\prime }}$ are the
basis components in the ''e'' coordinatization and thus not well-defined
tensor quantitities. The tensor quantity is $F^{ab}$ or the corresponding
quantities in the coordinate-based geometric language. Then we can write the
equalities for $F^{ab}$ in the same manner as in the relations for the
distance 4-vector (\ref{elt5}), i.e., that 
\begin{equation}
F^{ab}=F^{\alpha \beta }e_{\alpha }\otimes e_{\beta }=F^{\alpha ^{\prime
}\beta ^{\prime }}e_{\alpha ^{\prime }}\otimes e_{\beta ^{\prime
}}=F_{r}^{\alpha \beta }r_{\alpha }\otimes r_{\beta }=F_{r}^{\alpha ^{\prime
}\beta ^{\prime }}r_{\alpha ^{\prime }}\otimes r_{\beta ^{\prime }}.
\label{fa2}
\end{equation}
$F^{\alpha \beta }$ are the basis components, $e_{\alpha ,\beta }$ are the
basis 4-vectors, and $e_{\alpha }\otimes e_{\beta }$ is an outer product of
the basis 4-vectors, i.e., it is the basis for (2,0) tensors, and all
quantities are defined in $S$ and in the ''e'' coordinatization. Similarly
holds in the $S^{\prime }$ frame, and in the ''r'' coordinatization. The
basis components $F_{r}^{\alpha \beta },$ $F_{r}^{\alpha ^{\prime }\beta
^{\prime }},$ and the bases $r_{\alpha }\otimes r_{\beta },$ $r_{\alpha
^{\prime }}\otimes r_{\beta ^{\prime }}$ in the ''r'' cordinatization can be
found by applying the previously mentioned transformation matrix $T_{\quad
\nu ,r}^{\mu }$ to the corresponding quantities in the ''e''
coordinatization. \emph{The relations (\ref{fa2}) show that the basis
components }$F^{\alpha \beta }\emph{,}$\emph{\ }$F_{r}^{\alpha ^{\prime
}\beta ^{\prime }},$\emph{\ etc., when taken alone, are not equal. Only the
whole tensor quantity }$F^{ab}$\emph{\ and all the CBGQs }$F^{\alpha \beta
}e_{\alpha }\otimes e_{\beta },$\emph{\ }$F_{r}^{\alpha ^{\prime }\beta
^{\prime }}r_{\alpha ^{\prime }}\otimes r_{\beta ^{\prime }},$\emph{\ etc.,
are equal quantities, i.e., they are the same 4D physical quantity. }The
CBGQ $F^{\alpha ^{\prime }\beta ^{\prime }}e_{\alpha ^{\prime }}\otimes
e_{\beta ^{\prime }}$\ in $S^{\prime }$ and in the ''e'' coordinatization is
obtained by applying the Lorentz transformation $L^{\mu ^{\prime }}{}_{\nu
,e}$\ (\ref{lorus}) to the basis components $F^{\alpha \beta }$\ and the
inverse transformation $L^{\mu }{}_{\nu ^{\prime },e}$\ to the basis $%
e_{\alpha }\otimes e_{\beta }.$

However in the conventional derivation of the transformations for the
3-vectors $\mathbf{E}$ and $\mathbf{B,}$ (\ref{eitran}) and (\ref{eitr2}),
only \emph{some components in the specific coordinatization}, the ''e''
coordinatization, e.g., $F^{0i},$ of the whole tensor $F^{ab},$ i.e., of the 
$F^{\alpha \beta }e_{\alpha }\otimes e_{\beta }$, are taken alone (see (\ref
{eibi})), and they are argued to represent the components $E_{i}$ of some
''physical'' 3-vector $\mathbf{E.}$ As it is emphasized above, already this
step is meaningless in the ''TT relativity.'' We also remark that from the
mathematical point of view the identifications of the basis components $%
F^{0i}$ ($F^{0^{\prime }i^{\prime }}$) of $F^{ab}$ (that is of $F^{\alpha
\beta }e_{\alpha }\otimes e_{\beta }$ ($F^{\alpha ^{\prime }\beta ^{\prime
}}e_{\alpha ^{\prime }}\otimes e_{\beta ^{\prime }}$)) with the components $%
E_{i}$ ($E_{i^{\prime }}$) of the 3-vector $\mathbf{E}$ ($\mathbf{E}^{\prime
}$), the relations (\ref{eibi}) ((\ref{eib1})), are not permissible. Namely
it is not possible to say that, e.g., $E_{1}=F^{01},$ since the basis would
also need to be included, and the basis 3-vector $\mathbf{e}_{1}$ (in 3D
space) cannot be the same as the basis (2,0) tensor $e_{0}\otimes e_{1}.$
Further, as previously mentioned, the Lorentz transformation $L^{\mu
}{}_{\nu ,e}$\ (\ref{lorus}) transforms not only $F^{\alpha \beta }$ to $%
F^{\alpha ^{\prime }\beta ^{\prime }}$, but also the basis $e_{\alpha
}\otimes e_{\beta }$ to the basis $e_{\alpha ^{\prime }}\otimes e_{\beta
^{\prime }}$, whence it follows that neither in $S^{\prime }$ the components 
$E_{i^{\prime }}$ can be identified with the components $F^{0^{\prime
}i^{\prime }}.$ In fact, \emph{the basis components }$F^{0i},$\emph{\ or }$%
F^{0^{\prime }i^{\prime }},$\emph{\ when taken alone, are not well-defined
physical quantities in 4D spacetime. }

It is interesting to note that in the ''r'' cordinatization one finds that,
e.g., 
\begin{equation}
F_{r}^{01}=F^{01}+F^{12}+F^{13},\quad F_{r}^{0^{\prime }1^{\prime
}}=F^{01}-\beta \gamma (F^{02}+F^{03})+\gamma (F^{12}+F^{13})  \label{eref}
\end{equation}
which clearly shows that \emph{the identification of the components of the
the 3-vectors }$\mathbf{E}$\emph{\ and }$\mathbf{B}$\emph{\ with some of the
basis components of the tensor }$F^{ab}$\emph{\ is a coordinatization
dependent procedure and thus unphysical.}

However it is generally assumed in the usual approach (I am not aware of any
exception) that the two sets of basis components of the tensor $F^{ab}$ $%
\left\{ F^{0^{\prime }1^{\prime }},F^{0^{\prime }2^{\prime }},F^{0^{\prime
}3^{\prime }},F^{2^{\prime }3^{\prime }},F^{3^{\prime }1^{\prime
}},F^{1^{\prime }2^{\prime }}\right\} $ (i.e., the corresponding set of
components $\left\{ E_{1^{\prime }},E_{2^{\prime }},E_{3^{\prime
}},B_{1^{\prime }},B_{2^{\prime }},B_{3^{\prime }}\right\} $ obtained by the
identifications (\ref{eib1})) and $\left\{
F^{01},F^{02},F^{03},F^{23},F^{31},F^{12}\right\} $ (i.e., $\left\{
E_{1},E_{2},E_{3},B_{1},B_{2},B_{3}\right\} $ obtained by the
identifications (\ref{eibi})) \emph{represent the same quantity} for
observers in two relatively moving IFRs $S$ and $S^{\prime }.$ But the
relations (\ref{fa2}), and the above mentioned mathematical arguments,
explicitly show that the two mentioned sets of basis components of the
tensor $F^{ab}$ do not represent the same 4D quantity when considered in $S$%
\ and $S^{\prime }.$\ We can interpret these results saying that \emph{the
transformations (\ref{eitran}) (and (\ref{eitr2})) }that connect two
mentioned sets of basis components of $F^{ab}$ (i.e., that connect the
corresponding sets $\left\{ \mathbf{E}^{\prime },\mathbf{B}^{\prime
}\right\} $\ and $\left\{ \mathbf{E},\mathbf{B}\right\} $) actually \emph{%
connect different quantities in 4D spacetime}, and thus that \emph{they} 
\emph{are not the TT but the AT}.

It is worth noting that in many textbooks and papers on electrodynamics the
tensor $F^{ab}$ and the relations (\ref{fa2}) and (\ref{eref}) are never
mentioned, and almost always only the basis components (in the ''e''
coordinatization) $F^{\alpha \beta }$ of the tensor $F^{ab}$ are considered.
Moreover in almost every textbook or paper on electrodynamics $F^{\alpha
\beta }$ (the ''e'' coordinatization) is written as that its components are
the components of the 3-vectors $\mathbf{E}$ and $\mathbf{B}$ (see, e.g.,
Ref. 13 ch. 3, pp. 73,74, Ref. 10 Sec. 11.9). This means that $\mathbf{E}$
and $\mathbf{B}$ are considered as primary quantities while $F^{\alpha \beta
}$ is in some way a secondary quantity determined by the components of $%
\mathbf{E}$ and $\mathbf{B.}$ But, even if one works only in the ''e''
coordinatization, and considers only the covariant Maxwell equations \emph{%
for basis components} (\ref{maxco}) (thus not the true tensor equation with $%
F^{ab}$) it is more correct, from the ''TT relativity'' viewpoint, to take
that $F^{\alpha \beta }$ is the primary quantity; it is the solution of the
covariant Maxwell equations (\ref{maxco}), or the corresponding wave
equation $\partial ^{\sigma }\partial _{\sigma }F_{\alpha \beta
}-(1/\varepsilon _{0}c)(\partial _{\beta }j_{\alpha }-\partial _{\alpha
}j_{\beta })=0,$ and it conveys all the information about the
electromagnetic field. The solution $F^{\alpha \beta }$ (in the ''e''
coordinatization and in the retarded formulation) of these equations is
given as 
\[
F^{\alpha \beta }(x^{\mu })=(2k/i\pi c)\int \left\{ \frac{\left[ j^{\alpha
}(x^{\prime \mu })(x-x^{\prime })^{\beta }-j^{\beta }(x^{\prime \mu
})(x-x^{\prime })^{\alpha }\right] }{\left[ (x-x^{\prime })^{\sigma
}(x-x^{\prime })_{\sigma }\right] ^{2}}\right\} d^{4}x^{\prime } 
\]
where $x^{\alpha },x^{\prime \alpha }$ are the position 4-vectors of the
field point and the source point respectively, and $k=1/4\pi \varepsilon
_{0}.$ The corresponding CBGQ $F^{\alpha \beta }(x^{\mu })e_{\alpha }\otimes
e_{\beta }$ can be easily constructed and it is a well-defined quantity in
the ''TT relativity.'' We see that there is no need to introduce $E_{i}$ and 
$B_{i},$ and, in fact, \emph{to define }$F^{\alpha \beta }$\emph{\ by the
relations (\ref{eibi}) and (\ref{eib1})}, but one can work exclusively with
the components $F^{\alpha \beta }$ determined by the above expression for $%
F^{\alpha \beta }(x^{\mu })$.\bigskip \medskip

\noindent \textbf{5.3. Nonequivalence of the Maxwell Equations in the
3-Vector Form and in the Form of Tensor Equations }\medskip

The usual covariant Maxwell equations (\ref{maxco}) are actually the
equations in the ''e'' coordinatization for basis components in a chosen
IFR. We first show how these equations for the basis components are derived
from the true tensor equations (when no basis has been introduced). The true
tensor equations can be written in the abstract index notation as 
\begin{equation}
\nabla ^{a}F_{ab}=-j_{b}/\varepsilon _{0}c,\quad \varepsilon ^{abcd}\nabla
_{b}F_{cd}=0  \label{mxt3}
\end{equation}
where $\nabla _{b}$ is the derivative operator (sometimes called the
covariant derivative), see, e.g., Ref. 11. The tensor equation (\ref{mxt3})
can be written in the following form 
\begin{equation}
(-g)^{-1/2}\partial _{a}((-g)^{1/2}F^{ab})=-j^{b}/\varepsilon _{0}c,\quad
\varepsilon ^{abcd}\partial _{b}F_{cd}=0  \label{maxten}
\end{equation}
where $g$ is the determinant of the metric tensor $g_{ab}$ and $\partial
_{a} $ is an ordinary derivative operator. When some coordinatization is
chosen in an IFR $S$, e.g., the ''e'' coordinatization, then the relations (%
\ref{maxten}) can be written in the coordinate-based geometric language as
the equations that contain the basis vectors as well, 
\begin{equation}
\partial _{\alpha }F^{a\beta }e_{\beta }=-(1/\varepsilon _{0}c)j^{\beta
}e_{\beta },\quad \partial _{\alpha }\ ^{*}F^{\alpha \beta }e_{\beta }=0
\label{maco1}
\end{equation}
(We remark that (\ref{maco1}) follows from (\ref{maxten}) for those
coordinatizations for which the basis vectors are constant, e.g., ''e'' and
''r'' coordinatizations considered here. For a nonconstant basis, for
example, when one uses polar or spherical basis one forms and vectors (and
''e'' or ''r'' synchronization) then one must also differentiate these
nonconstant basis vectors.) From (\ref{maco1}), which contain the basis
(1,0) tensors (4-vectors), one finds the already written equations for basis
components (\ref{maxco}); every equation in (\ref{maco1}) is the equality of
two tensors of the same type, two 4-vectors, and if two 4-vectors are equal
then the corresponding components are equal, and that holds in all bases. In
many treatments only the covariant Maxwell equations (\ref{maxco}) for the
basis components are used forgetting that they are obtained from the tensor
equations (\ref{maxten}) or (\ref{maco1}). Then, with substitutions (\ref
{eibi}), one finds the equations for the quantities $E_{i}$ and $B_{i}$ 
\begin{eqnarray}
\partial _{i}E_{i} &=&\rho /\varepsilon _{0},\quad \varepsilon
_{ijk}\partial _{j}E_{k}=-c\partial _{0}B_{i}  \nonumber \\
\partial _{i}B_{i} &=&0,\quad \varepsilon _{ijk}\partial _{j}B_{k}=\mu
_{0}j_{i}+\mu _{0}\varepsilon _{0}c\partial _{0}E_{i}  \label{eimax}
\end{eqnarray}
>From the quantities $E_{i}$ and $B_{i}$ \emph{and the basis 3-vectors in 3D
space }$\mathbf{e}_{i}$ one constructs the electric and magnetic fields
3-vectors as $\mathbf{E=}E_{i}\mathbf{e}_{i},$ $\mathbf{B}=B_{i}\mathbf{e}%
_{i}$ in $S.$ Then the equations for the quantities $E_{i}$ and $B_{i}$ (\ref
{eimax}) are written as the usual Maxwell equations for the 3-vectors $%
\mathbf{E}$ and $\mathbf{B}$ (\ref{max}). This is obviously a very awkward
procedure; we started from the equation (\ref{maco1}) with the basis vectors
in 4D spacetime and ended in the equation (\ref{max}) with the basis vectors
in 3D space. One can form an equation with the components of tensors if
these components are combined only by the permissible tensor operations. The
usual Maxwell equations (\ref{eimax}) are not obtained by such permissible
tensor operations with the components of tensors since the identifications (%
\ref{eibi}) are not such operations. Furthermore the Maxwell equations (\ref
{max}) contain the 3-vectors $\mathbf{E}$ and $\mathbf{B,}$ and these
3-vectors are constructed in an artifical way from \emph{some basis
components of the tensor }$F^{\alpha \beta }e_{\alpha }\otimes e_{\beta }$%
\emph{\ and the basis 3-vectors }$\mathbf{e}_{i}$ of the 3D space. \emph{The
matematical equivalence between the equations with the basis 3-vectors and
the equations with the basis 4-vectors cannot exist. Thence we can conclude
that the usual Maxwell equations with }$E_{i}$\emph{\ and }$B_{i}$\emph{\ (%
\ref{eimax}), or with the 3-vectors }$\mathbf{E}$\emph{\ and }$\mathbf{B}$%
\emph{\ (\ref{max}), are not equivalent to the tensor equations (\ref{maxten}%
), i.e., to the CBGEs (\ref{maco1}).}

If one uses the ''r'' cordinatization then from the tensor equation (\ref
{maxten}) one finds a similar equation to (\ref{maco1}) but now in the ''r''
coordinatization, 
\begin{equation}
\partial _{\alpha r}F_{r}^{a\beta }e_{\beta r}=-(1/\varepsilon
_{0}c)j_{r}^{\beta }e_{\beta r},\quad \partial _{\alpha r}\
^{*}F_{r}^{\alpha \beta }e_{\beta r}=0  \label{maxer}
\end{equation}
Note that in IFRs $g$ is constant and $g=-1$. From these equations (\ref
{maxer}), which contain the basis vectors, and \emph{which are completely
equivalent in the description of physical phenomena to the corresponding
equations in the ''e'' coordinatization }(\ref{maco1}), one naturally finds
the equations for basis components 
\begin{equation}
\partial _{\alpha r}F_{r}^{a\beta }=-(1/\varepsilon _{0}c)j_{r}^{\beta
},\quad \partial _{\alpha r}\ ^{*}F_{r}^{\alpha \beta }=0  \label{ercomp}
\end{equation}
Let us perform the similar identifications of some of the basis components
of $F^{ab}$ with the quantities $E_{ir}$ and $B_{ir},$ corresponding in the
''r'' coordinatization to the quantities $E_{i}$ and $B_{i}$ in the ''e''
coordinatization. Then, as can be seen from the relations (\ref{eref}), one 
\emph{will not get from (\ref{ercomp}) the equations of the same form as the
corresponding equations in the ''e'' coordinatization} (\ref{eimax}). This
is caused by the fact that the identifications of some of the basis
components of the tensor $F^{ab}$ with the components of the 3-vectors are
not the permissible tensor operations. Of course, if one introduces the
basis 3-vectors $\mathbf{r}_{i}$ in the ''r'' coordinatization in 3D space,
and constructs the 3-vectors $\mathbf{E}_{r}\mathbf{=}E_{ir}\mathbf{r}_{i}$
and $\mathbf{B}_{r}=B_{ir}\mathbf{r}_{i},$ then the 3-vectors in the ''r''
coordinatization and in the ''e'' coordinatization are not equal, i.e., $%
\mathbf{E}_{r}\neq \mathbf{E}$ and $\mathbf{B}_{r}\neq \mathbf{B.}$ The
obtained equations with $\mathbf{E}_{r}$ and $\mathbf{B}_{r}$ are not of the
same form as the usual Maxwell equations with $\mathbf{E}$ and $\mathbf{B}$ (%
\ref{max}). This additionally proves that the Maxwell equations with the
3-vectors are not equivalent to the Maxwell equations in tensor form (\ref
{maxten}), or (\ref{maco1}), or (\ref{maxer}).

In the $\left\{ e_{\mu ^{\prime }}\right\} $ basis the relation (\ref{maxten}%
) becomes 
\begin{equation}
\partial _{\alpha ^{\prime }}F^{a^{\prime }\beta ^{\prime }}e_{\beta
^{\prime }}=-(1/\varepsilon _{0}c)j^{\beta ^{\prime }}e_{\beta ^{\prime
}},\quad \partial _{\alpha ^{\prime }}\ ^{*}F^{\alpha ^{\prime }\beta
^{\prime }}e_{\beta ^{\prime }}=0  \label{maxc2}
\end{equation}
which then gives \emph{the equation for basis components} (\ref{maxc1}).
Again, by the same reasoning as above (after the relation (\ref{eimax})), we
conclude that the usual Maxwell equations in the $S^{\prime }$ frame, i.e.,
with $E_{i^{\prime }}$ and $B_{i^{\prime }}$, or with the 3-vectors $\mathbf{%
E}^{\prime }$ and $\mathbf{B}^{\prime }$, are not equivalent to the tensor
equations (\ref{maxten}), that is, to the CBGEs (\ref{maxc2}).

In addition to the comparison of the tensor Maxwell equations (the CBGEs)
with the Maxwell equations in the 3-vector form we also investigate the
relation of the tensor Maxwell equations and the equations with the basis
components. Some important conclusions will be derived regarding the
mathematical form of the physical laws in the ''TT relativity.'' From the
mathematical viewpoint the (1,0) tensor quantity $(-g)^{-1/2}\partial
_{a}((-g)^{1/2}F^{ab})$ can be written in the coordinate-based geometric
language in the ''e'' cordinatization, and in $S$ as $\partial _{\alpha
}F^{a\beta }e_{\beta },$ while in $S^{\prime }$ as $\partial _{\alpha
^{\prime }}F^{a^{\prime }\beta ^{\prime }}e_{\beta ^{\prime }},$ where all
primed quantities (including the basis vectors) are obtained by the TT,
i.e., by the Lorentz transformation $L^{\mu }{}_{\nu ,e}$ (\ref{lorus}) from
the corresponding unprimed quantities. The same holds in the ''r''
coordinatization; the quantities in the ''r'' coordinatization can be
determined from the corresponding quantities in the ''e'' coordinatization
using the previously mentioned transformation matrix $T_{\quad \nu ,r}^{\mu
} $, which is also a TT. Thus 
\begin{equation}
(-g)^{-1/2}\partial _{a}((-g)^{1/2}F^{ab})=\partial _{\alpha }F^{a\beta
}e_{\beta }=\partial _{\alpha ^{\prime }}F^{a^{\prime }\beta ^{\prime
}}e_{\beta ^{\prime }}=\partial _{\alpha r}F_{r}^{a\beta }e_{\beta
r}=\partial _{\alpha ^{\prime }r}F_{r}^{a^{\prime }\beta ^{\prime }}e_{\beta
^{\prime }r}  \label{mxj}
\end{equation}
which shows that \emph{the equalities in (\ref{mxj}) refer to the same
quantity in 4-D spacetime.} Analogously, the mathematics yields for the
(1,0) tensor (4-vector) $-j^{b}/\varepsilon _{0}c$ the relations 
\begin{equation}
-j^{b}/\varepsilon _{0}c=-(1/\varepsilon _{0}c)j^{\beta }e_{\beta
}=-(1/\varepsilon _{0}c)j^{\beta ^{\prime }}e_{\beta ^{\prime
}}=-(1/\varepsilon _{0}c)j_{r}^{\beta }e_{\beta r}=-(1/\varepsilon
_{0}c)j_{r}^{\beta ^{\prime }}e_{\beta ^{\prime }r}  \label{mx2}
\end{equation}
A similar analysis can be applied to the second Maxwell equation in (\ref
{maxten}). \emph{The physical laws} expressed as tensor equations, e.g., (%
\ref{maxten}), or equivalently as CBGEs, for example, (\ref{maco1}), (\ref
{maxc2}) and (\ref{maxer}),\emph{\ set up the connection between two
geometric quantities}, in this case, two 4-vectors, that are given by
equations (\ref{mxj}) and (\ref{mx2}). \emph{The experiments in which all
parts of tensor quantities are measured then play the fundamental role in
deciding about the validity of some physical law mathematically expressed as
tensor equation. }We see from the equations (\ref{mxj}) and (\ref{mx2}) that 
\emph{when the physical laws are expressed as tensor, geometric, equations} (%
\ref{maxten}), (\ref{maco1}), (\ref{maxc2}) and (\ref{maxer}), (or, in other
words, when the transformations that connect relatively moving IFRs, and
different coordinatizations of the chosen IFR, are all the TT), \emph{then
these equations are invariant upon the Lorentz transformations and the
transformations} $T_{\ \nu }^{\mu }$ (\ref{lamb}) between different
coordinatizations. It is not so for the equations in the component form,
e.g., (\ref{maxco}), (\ref{maxc1}), (\ref{ercomp}). First, one cannot write
such equalities, (\ref{mxj}) and (\ref{mx2}), for the basis components
alone. Although, for example, the quantity $\partial _{\alpha }F^{a\beta }$
from the equations for the basis components (\ref{maxco}) is of \emph{the
same form} as $\partial _{\alpha ^{\prime }}F^{a^{\prime }\beta ^{\prime }}$
from (\ref{maxc1}), with primed quantities replacing the unprimed ones,
these quantities $\partial _{\alpha }F^{a\beta }$\ and $\partial _{\alpha
^{\prime }}F^{a^{\prime }\beta ^{\prime }}$\ are not equal.\emph{\ }When
they are taken without the basis vectors then they are not the same 4D
quantity. Further, one sees that \emph{for the equations with the basis
components only the form of equations remains unchanged under the Lorentz
transformations and the transformations }$T_{\ \nu }^{\mu }$ (\ref{lamb}),
i.e., \emph{such equations are covariant but not invariant}. Of course the
covariance of physical equations, when they are written in the component
form, is a simple consequence of the invariance of tensor quantities, or
equivalently, of the CBGQs, upon the mentioned TT, that is upon the
isometries. \emph{The invariance of physical laws, that are expressed as
tensor equations, or equivalently as the CBGEs, means that all physical
phenomena proceed in the same way (taking into account the corresponding
initial and boundary conditions) in different IFRs. Thus there is no
physical difference between these frames, what automatically embodies the
principle of relativity. We remark that in the ''TT relativity'' there is no
need to postulate the principle of relativity as a fundamental law. It is
replaced by the requirement that the physical laws must be expressed as
tensor equations (or equivalently as the CBGEs) in the 4D spacetime.}

The above consideration shows that the Maxwell equations in the 3-vector
form are not equivalent to the Maxwell equations when written as tensor
equations, or as CBGEs. But, in addition, we explicitly show that the
Maxwell equations in the 3-vector form are not the covariant equations,
i.e., that their \emph{form }do not remain unchanged when they are
transformed by the Lorentz transformation from $S$ to $S^{\prime }$. It is
the consequence of the fact that the transformations of the 3-vectors $%
\mathbf{E}$ and $\mathbf{B,}$ the relations (\ref{eitr2}), are the AT, which
do not refer to the same 4D quantities. When the Lorentz transformations (%
\ref{lorus}), as the TT (in 4D spacetime), are applied to the basis
components $x^{\mu },$ $\partial _{\mu }$ and $j^{\mu }$ in the usual
Maxwell equations for the quantities $E_{i}$ and $B_{i}$ (\ref{eimax}) and
the AT (\ref{eitran}) are applied to $E_{i}$ and $B_{i}$ then we find 
\begin{eqnarray}
\partial _{i^{\prime }}E_{i^{\prime }} &=&\rho ^{\prime }/\varepsilon
_{0}-V_{i}\left[ \varepsilon _{ijk}\partial _{j^{\prime }}B_{k^{\prime
}}-\mu _{0}j_{i^{\prime }}-\mu _{0}\varepsilon _{0}c\partial _{0^{\prime
}}E_{i^{\prime }}\right]  \nonumber \\
\varepsilon _{ijk}\partial _{j^{\prime }}E_{k^{\prime }} &=&-c\partial
_{0^{\prime }}B_{i^{\prime }}+V_{i}\left[ \partial _{j^{\prime
}}B_{j^{\prime }}\right]  \nonumber \\
\partial _{i^{\prime }}B_{i^{\prime }} &=&(V_{i}/c^{2})\left[ \varepsilon
_{ijk}\partial _{j^{\prime }}E_{k^{\prime }}+c\partial _{0^{\prime
}}B_{i^{\prime }}\right]  \label{eima2} \\
\varepsilon _{ijk}\partial _{j^{\prime }}B_{k^{\prime }} &=&\mu
_{0}j_{i^{\prime }}+\mu _{0}\varepsilon _{0}c\partial _{0^{\prime
}}E_{i^{\prime }}-(V_{i}/c^{2})\left[ \partial _{i^{\prime }}E_{i^{\prime
}}-\rho ^{\prime }/\varepsilon _{0}\right]  \nonumber
\end{eqnarray}
Obviously the Maxwell equations (\ref{eima2}) in $S^{\prime }$ do not have
the same form as in $S$ (\ref{eimax}). Note that if the fourth relation from
(\ref{eima2}) was introduced into the first relation from (\ref{eima2}) then
one obtains $(\partial _{i^{\prime }}E_{i^{\prime }}-\rho ^{\prime
}/\varepsilon _{0})(1-V^{2}/c^{2})=0$ and similarly for other relations. For 
$V\neq c$ one finally finds the same form for the Gauss law in $S^{\prime }$
as it is in $S,$ i.e., $\partial _{i^{\prime }}E_{i^{\prime }}-\rho ^{\prime
}/\varepsilon _{0}=0.$ But different manipulations were needed to achieve
the same form of laws in $S^{\prime }$ as in $S.$ (The result (\ref{eima2})
has been already presented in Ref. 17.)

The same result as (\ref{eima2}) was mentioned in Ref. 13 Sec.3.4. but there
it was obtained and interpreted in a different way. We discuss that
derivation in order to show some important differences between the ''TT
relativity'' and the usual approach to special relativity. In Ref. 13 they
start with the Gauss law for the magnetic field in the 3-vector form (all is
done in the ''e'' coordinatization) in a specific IFR $S,$ i.e., $\nabla 
\mathbf{B}=\partial _{i}B_{i}=0.$ Then the principle of relativity is used
in a way that is usual in the conventional ''AT relativity.'' They asserts
that the statement $\nabla \mathbf{B}=0:$ ''has to be true in all Lorentz
frames,'' and hence they write this equation in some relatively moving IFR $%
S^{\prime }$ as $\partial _{i^{\prime }}B_{i^{\prime }}=0.$ In such usual
approach the principle of relativity is understood as a fundamental
postulate according to which any physical law ''has to be true in all
Lorentz frames,'' and \emph{the Gauss law in the 3-vector form }$\nabla 
\mathbf{B}=0$ is considered to be a physical law. Note that such an
understanding of the principle of relativity does not exist in the ''TT
relativity.'' There, as shown above, it is required that the physical laws
must be expressed as tensor equations (or equivalently as the CBGEs) in the
4D spacetime. The invariance of physical laws upon the TT automatically
follows from such formulation. \emph{The relations between the 3-vectors
cannot be the physical laws that hold in 4D spacetime, since such relations
will necessarily change their form upon the Lorentz transformation, which
are the transformations defined on 4D spacetime. }Let us see how such
changes necessarily appear in the considered derivation in Ref. 13 Sec.3.4.,
and how they are interpreted. The authors of$^{(13)}$ continue the
derivation substituting the usual transformations of $E_{i^{\prime }}$\ and $%
B_{i^{\prime }}$\ (\ref{eitran}) and the Lorentz transformations of $%
\partial _{\mu ^{\prime }}\equiv \partial /\partial x^{\mu ^{\prime }}$ into
the Gauss law $\partial _{i^{\prime }}B_{i^{\prime }}=0$ in $S^{\prime }$.
(In fact, they simplified the derivation taking that $\beta \ll 1,$ i.e.,
that $\gamma =1.$ However such simplification is unnecessary and the
complete Lorentz transformations and the transformations (\ref{eitran}) lead
to the same result.) Now comes an interesting point which nicely illustrates
how some problems, obviously appearing in the approach with the 3-vectors,
are artificially avoided and even wrongly interpreted. After the mentioned
substitutions it is stated in Ref. 13: ''Recover the original condition of
zero divergence in the laboratory frame, plus the following additional
information (requirement for the vanishing of the coefficient of the
arbitrary small velocity $\beta $): 
\[
\partial B_{x}/\partial t+\partial E_{z}/\partial y-\partial E_{y}/\partial
z=0" 
\]
Thus they started with $\partial _{i^{\prime }}B_{i^{\prime }}=0,$ then
performed the transformations and finally they obtained the following
relation $\partial _{i}B_{i}-(V/c^{2})\left[ \varepsilon _{1jk}\partial
_{j}E_{k}+c\partial _{0}B_{1}\right] =0,$ since they chose that the velocity
of transformation been directed in the $x-$ direction. This is exactly our
relation (\ref{eima2}) (the third equation, but for the reversed
transformation). We derived (\ref{eima2}) using the Lorentz transformations (%
\ref{lorus}) and the transformations of $E_{i^{\prime }}$\ and $B_{i^{\prime
}}$\ (\ref{eitran}), which hold not only for $V\ll c$ but for any $V\prec c.$
Hence it follows that it is not possible to put into the equation $\partial
_{i}B_{i}-(V/c^{2})\left[ \varepsilon _{1jk}\partial _{j}E_{k}+c\partial
_{0}B_{1}\right] =0$ that $\varepsilon _{1jk}\partial _{j}E_{k}+c\partial
_{0}B_{1}=0$ as done in Ref. 13 ''(requirement for the vanishing of the
coefficient of the arbitrary small velocity $\beta $)$;$'' the velocity does
not need to be arbitrary small. Simply they obtained the contradiction; they
started from the assertion that the Gauss law for the magnetic field in the
3-vector form must have the same form in all Lorentz frames (the generally
accepted formulation of the principle of relativity) and found by applying
the Lorentz transformations and the usual transformations of $E_{i}$\ and $%
B_{i}$\ (\ref{eitran}) that it is not true. The above discussion in this
section reveals that there are two reasons for the changes of form of the
usual Maxwell equations in the 3-vector form. The first reason is the
traditional formulation of the principle of relativity in which this
principle acts as the postulate established outside the mathematical
formulation of the theory. In contrast to this in the ''TT relativity'' the
equivalence of all Lorentz frames in the description of the physical
phenomena naturally follows, as already said, from the formulation of
physical laws as tensor equations, or equivalently as the CBGEs. The second
reason is the use of the ''apparent transformations'' of $E_{i}$\ and $B_{i}$%
\ (\ref{eitran}).

When Einstein$^{(1)}$ derived the transformations of the 3-vectors $\mathbf{E%
}$ and $\mathbf{B,}$ the relations (\ref{eitr2}) or (\ref{eitran}), he made
it using the principle of relativity as a postulate and using the Maxwell
equations in the 3-vector form. Let us discuss his derivation of (\ref{eitr2}%
). Einstein worked with the Maxwell equations (\ref{max}) (thus in the ''e''
coordinatization) and first he performed the Lorentz transformations of the
derivatives from an IFR $S$ (he denoted it as $K$) to another IFR $S^{\prime
}$ ($k$ in his notation) moving with $V$ relative to $S$. In the course of
that derivation he combined different Maxwell equations and, for example, he
found $\partial _{0^{\prime }}E_{i}$ from the Amp\`{e}re-Maxwell law, the
last relation in (\ref{max}), and inserted it into the Gauss law for $%
\mathbf{E}$, the first relation in (\ref{max}). This is an important step in
the derivation. But, strictly speaking, such a combination of the
transformation relations for different equations (physical laws) is not, in
fact, allowed, since one first has to know how every law (separately) will
look like in another IFR. Then, after regrouping different terms, e.g., in
Gauss's law, so that the equations have the same form in $S^{\prime }$ as in
the original frame $S$, he used the principle of relativity, Ref. 1 p.52:
''Now the principle of relativity requires that if the Maxwell-Hertz
equations for empty space hold good in system $K$, they also hold good in
system $k$, .. .'' This led him to the relations (\ref{eitran}) for $E_{i}$\
and $B_{i}$. But, as we see from (\ref{eima2}) and from the discussion of
the derivation given in Ref. 13, the Maxwell equations in the 3-vector form
do not remain unchanged when the Lorentz transformations (\ref{lorus}) are
applied to the derivatives $\partial _{\mu }$ and the transformations (\ref
{eitran}) are applied to $E_{i}$ and $B_{i}.$ Hence the same objections and
remarks hold for this Einstein's derivation of (\ref{eitran}) as we
considered above for Ref. 13.

We can conclude from the whole discussion that the Maxwell equations in the
3-vector form (\ref{max}), or equivalently (\ref{eimax}) (always only in the
''e'' coordinatization), are neither covariant (they change their form when
going from an IFR $S$ to another relatively moving IFR $S^{\prime }$), nor
are they equivalent to the tensor Maxwell equations (\ref{maxten}), or to
the CBGEs (\ref{maco1}), (\ref{maxc2}), (\ref{maxer}).

It is also remarked in Ref. 20 that the 3-vector form of the Maxwell
equations in a noninertial frame is not unique, but it depends on which
components of the electromagnetic field tensor (contravariant, covariant,
mixed) one uses to identify them with the components of the electric and
magnetic 3-vectors. In a comment$^{(21)}$ on$^{(20)}$ it is explained that
there is not any real ambiguity in defining electric and magnetic 3-vector
fields in noninertial and curved spacetimes if one introduces the electric $%
E^{a}$ and magnetic $B^{a}$ 4-vector fields instead of the usual 3-vector
fields. Our consideration revealed that \emph{in IFRs the Maxwell equations
in the 3-vector form (\ref{max}) are not equivalent to the tensor Maxwell
equations (\ref{maxten}), which means that the introduction of the electric }%
$E^{a}$\emph{\ and magnetic }$B^{a}$\emph{\ 4-vector fields is necessary in
IFRs as well.} The covariant formulation of vacuum electrodynamics in IFRs
with the basis components of the 4-vectors $E^{a}$ and $B^{a}$ is considered
in Refs. 2 and 3 and here we shall derive some important results in a more
general manner.

Before doing it we consider another derivation of (\ref{eitran}) or (\ref
{eitr2}), which is presented in, e.g., the well-known textbook,$^{(18)}$ and
which explicitly uses the Lorentz ''contraction'' (\ref{apcon}). From the
''TT relativity'' viewpoint such derivation directly shows that the
transformations of the 3-vectors $\mathbf{E}$ and $\mathbf{B}$ (\ref{eitr2})
(or (\ref{eitran})) are the AT. Here we discuss some important steps and
results in Purcell's derivation. First Purcell derives the expressions for
the charge and current densities (for example, the relations (53) and (54)
in Sec. 6 in Ref. 18) assuming that \emph{the special relativity requires
the Lorentz ''contraction,''} e.g., of the moving charge sheets in Fig. 5.9,
or of the distance between moving positive ions in Fig. 5.20 in Ref. 18.
Then he implicitly deals with the conventional definition of charge 
\[
Q=(1/c)\int_{V(t)}j^{0}(\mathbf{r},t)dV 
\]
e.g., when determining the charge density for charge sheet in Fig. 5.9. In
this definition the volume $V(t)$ is taken at a particular time $t$ and it
is stationary in some IFR $F.$ The values of the charge density $\rho (%
\mathbf{r},t)=j^{0}(\mathbf{r},t)/c$ are taken simultaneously for all $%
\mathbf{r}$ in $V(t).$ It is supposed in the usual tretments (including$%
^{(18)}$) that the volume elements $dV^{\prime }$ are Lorentz ''contracted''
in a relatively moving IFR $F^{\prime }$ and all of them, i.e., the whole
volume $V^{\prime }(t^{\prime }),$ are taken simultaneously at some
arbitrary $t^{\prime }$ in $F^{\prime }$. $t^{\prime }$\emph{\ in }$%
F^{\prime }$\emph{\ is not connected in any way with }$t$\emph{\ in }$F$.
Furthermore it is assumed that $j^{0}$ from $F$ is transformed (using the
Lorentz ''contraction'') only to $j^{0^{\prime }}$ in $F^{\prime }$ and all $%
j^{0^{\prime }}$ are taken simultaneously at the same $t^{\prime }$ in $%
F^{\prime }$. The new $Q^{\prime }=(1/c)\int_{V^{\prime }(t^{\prime
})}j^{0^{\prime }}(\mathbf{r}^{\prime },t^{\prime })dV^{\prime }$ in $%
F^{\prime }$ is considered to be equal to the charge $Q$ in $F,$ $Q^{\prime
}=Q$ (the total charge is invariant). But we remark that the charge $Q$
defined in such a manner cannot be invariant upon the Lorentz transformation 
$L^{\mu ^{\prime }}{}_{\nu ,e}.$ As shown in Sec. 4.1 the Lorentz
''contraction'' has nothing to do with the Lorentz transformation, and the
Lorentz transformation $L^{\mu ^{\prime }}{}_{\nu ,e}$ cannot transform one
component $j^{0}$ from an IFR $F$ to the same component $j^{0^{\prime }}$ in 
$F^{\prime }.$ Also if all $j^{0}$ values are taken simultaneously at some $%
t $ in $F$ then the Lorentz transformation $L^{\mu ^{\prime }}{}_{\nu ,e}$
cannot transform them to the values $j^{0^{\prime }}$ which are all
simultaneous at some arbitrary $t^{\prime }$ in $F^{\prime }$. This
consideration shows that such an ''AT relativity'' definition of charge
cannot be relativistically correct definition. In order to connect the
charge densities with the electric fields (3-vectors) Purcell introduces
another definition of charge in terms of the Gauss law (for the 3-vector $%
\mathbf{E}$) when written in the integral form 
\[
Q=\varepsilon _{0}\int_{S(t)}\mathbf{E}(\mathbf{r},t)d\mathbf{a} 
\]
Eq. (3) Sec. 5 in Ref. 18. The whole above discussion and all objections to
the definition of charge by means of the volume integral of the charge
density exactly apply to this definition of $Q$ in terms of the flux of the
3-vector $\mathbf{E}$. In fact, the same objections hold also for \emph{all} 
\emph{Maxwell's equations in the 3-vector form }when they are written \emph{%
in the integral form}. Further it is formulated in Ref. 18 ''a formal
statement of the relativistic invariance of charge'' by Eq. (4) Sec. 5 
\[
\int_{S(t)}\mathbf{E}(\mathbf{r},t)d\mathbf{a=}\int_{S^{\prime }(t^{\prime
})}\mathbf{E}^{\prime }(\mathbf{r}^{\prime },t^{\prime })d\mathbf{a}^{\prime
} 
\]
and it is claimed there that: ''Each of the surface integrals in Eq. 4 is to
be evaluated at one instant in \emph{its} frame.'' Again we note that $%
t^{\prime },$ $S^{\prime },$ and $d\mathbf{a}^{\prime }$ in the moving frame 
$F^{\prime }$ are not obtained by the Lorentz transformation $L^{\mu
^{\prime }}{}_{\nu ,e}$ from the corresponding quantities in the frame $F$.
Actually, as said above, $t^{\prime }$ is an arbitrary time in $F^{\prime }$
and the surface $S^{\prime }$ is considered to be obtained from the surface $%
S$ by the Lorentz ''contraction,'' see, e.g., Sec. 5.5 in Ref. 18. Thus the
''AT relativity'' definition of a charge $Q$ by the flux of $\mathbf{E}$
cannot be relativistically correct definition either. Consequently the above
equality of fluxes of $\mathbf{E}$ in $F$ and $F^{\prime }$ (Eq. (4) Sec. 5
in Ref. 18) has nothing to do with the relativistic invariance of charge
upon the Lorentz transformation. In addition we mention that Purcell, in the
same way as many others, treats the transformation of the force as the AT
and not as the TT (in this case the Lorentz transformation). Namely, as can
be seen from Eq. (23) Sec.5 in Ref. 18, \emph{only some spatial components}
of the 4-force are compared in $F$ and $F^{\prime }$ and not the whole
4-vector. Furthermore the expressions for the fields $\mathbf{E}$ and $%
\mathbf{B,}$ e.g., Eqs. (55) and (56) Sec. 6 in Ref. 18, are determined
invoking the postulate of relativity. But that postulate is understood in
the same sense as we discussed above in connection with Einstein's
derivation of (\ref{eitr2}) and in the mentioned example from,$^{(13)}$
i.e., in a typical ''AT relativity'' manner. Since in Ref. 18 the
transformations of $\mathbf{E}$ and $\mathbf{B}$ are obtained by using the
AT of different quantities, and also using different definitions which
contain such transformations, it is clear that the transformations (\ref
{eitr2}) (or (\ref{eitran})) are the AT as well.

The whole procedure and the results can be interpreted from the ''TT
relativity'' viewpoint. The current-density 4-vector $j^{\mu ^{\prime
}}e_{\mu ^{\prime }}$ in the moving frame (all is in the ''e''
coordinatization) has to be determined from the $j^{\mu }e_{\mu }$ 4-vector
in the rest frame of the charges, in a similar manner as in Ref. 3 (but
there the quantities were written only in the component form). Thus in the
''TT relativity'' approach one has to use the Lorentz transformation $L^{\mu
^{\prime }}{}_{\nu ,e}$ instead of the Lorentz ''contraction''. Further, the
total electric charge $Q$ in a three-dimensional hypersurface $H$ (with
two-dimensional boundary $\delta H$) is defined by the tensor equation 
\begin{equation}
Q_{\delta H}=(1/c)\int_{H}j^{a}t_{a}dH  \label{charge}
\end{equation}
where $t_{a}$ is the unit normal to $H$, see, e.g., Ref. 11 ch. 4 and Ref.
3. The Gauss law in the integral form can be also written as the tensor
equation using the electric 4-vector field $E^{a}$ (see the next section,
Eq. (\ref{veef}) ) as 
\begin{equation}
(1/c)\int_{H}j^{a}t_{a}dH=\varepsilon _{0}\int_{_{\delta H}}E^{a}n_{a}dA
\label{gaus}
\end{equation}
where the integral on the right hand side of (\ref{gaus}) is the integral of
the normal component of $E^{a}$ on $\delta H$, see Ref. 11. (Wald uses such
form of Gauss's law in curved spacetime but our discussion shows that such
form has to be used in flat spacetime, and particularly in IFRs, as well.)
The relativistic invariance of charge automatically follows from the
definition (\ref{charge}) or (\ref{gaus}). In an ''invariant'' approach to
the SR, i.e., in the ''TT relativity,'' the definitions (\ref{charge}) and (%
\ref{gaus}) replace the above quoted usual definitions of $Q$ in terms of
the volume integral of charge density and by means of the flux of the
3-vector $\mathbf{E.}$ 
\bigskip \medskip

\noindent \textbf{6. COVARIANT\ ELECTRODINAMICS\ WITH\ }$E^{a}$\textbf{\ and }$B^{a}$
\bigskip

In accordance with the discussion in the previous section we introduce the
4-vectors $E^{a}$ and $B^{a}$ (see also Refs. 11, 21 and 22) instead of the
usual 3-vectors $\mathbf{E}$ and $\mathbf{B,}$ in order to formulate the
Maxwell equations as tensor equations with $E^{a}$ and $B^{a},$ which will
be equivalent to the tensor Maxwell equations (\ref{mxt3}), (\ref{maxten}),
with $F^{ab}.$ Then we define the electric and magnetic field by the
relations 
\begin{equation}
E_{a}=(1/c)F_{ab}v^{b},\quad B^{a}=-(1/2c^{2})\varepsilon ^{abcd}v_{b}F_{cd}
\label{veef}
\end{equation}
The $E^{a}$ and $B^{a}$ are the electric and magnetic field 4-vectors
measured by an observer moving with 4-velocity $v^{a}$ in an arbitrary
reference frame, $v^{a}v_{a}=-c^{2},$ and $\varepsilon ^{abcd}$ is the
totally skew-symmetric Levi-Civita pseudotensor (density). These fields
satisfy the conditions $v_{a}E^{a}=v_{b}B^{b},$ which follow from the
definitions (\ref{veef}) and the antisymmetry of $F_{ab}.$ In the usual
treatments (see, e.g., Refs. 11, 22 and 21) the tensors $E^{a}$ and $B^{a}$
are introduced in the curved spacetimes or noninertial frames, but at the
same time the usual Maxwell equations with the 3-vectors $\mathbf{E}$ and $%
\mathbf{B}$ are considered to be valid in the IFRs. One gets the impression
that $E^{a}$ and $B^{a}$ are considered only as useful mathematical objects,
while the real physical meaning is associated with the 3-vectors $\mathbf{E}$
and $\mathbf{B.}$ Our results obtained in the preceding sections imply that
it is necessary to use the 4-vectors $E^{a}$ and $B^{a}$ in IFRs as well.
This means that the tensor quantities $E^{a}$ and $B^{a}$ do have the real
physical meaning and not the 3-vectors $\mathbf{E}$ and $\mathbf{B.}$ The
inverse relation connecting the $E^{a},B^{a}$ and the tensor $F_{ab}$ is 
\begin{equation}
F_{ab}=(1/c)(v_{a}E_{b}-v_{b}E_{a})+\varepsilon _{abcd}v_{c}B_{d}\ 
\label{vezFE}
\end{equation}
The tensor Maxwell equations with $E^{a},B^{a}$ in the curved spacetimes are
derived in Ref. 22. Here we specify them to the IFRs, but in such a way that
they remain valid for different coordinatizations of the chosen IFR. First
we write the tensor Maxwell equations (\ref{maxten}) with $F^{ab}$ as the
CBGEs (\ref{maco1}) $\partial _{\alpha }F^{a\beta }e_{\beta
}=-(1/\varepsilon _{0}c)j^{\beta }e_{\beta },\quad \partial _{\alpha }\
^{*}F^{\alpha \beta }e_{\beta }=0.$ Then we also write the equation (\ref
{vezFE}) in the coordinate-based geometric language and the obtained
equation substitute into (\ref{maco1}) (all is done in the ''e''
coordinatization). This procedure yields 
\begin{eqnarray}
\partial _{\alpha }(\delta _{\quad \mu \nu }^{\alpha \beta }v^{\mu }E^{\nu
}+c\varepsilon ^{\alpha \beta \mu \nu }B_{\mu }v_{\nu })e_{\beta }
&=&-(j^{\beta }/\varepsilon _{0})e_{\beta }  \nonumber \\
\partial _{\alpha }(\delta _{\quad \mu \nu }^{\alpha \beta }v^{\mu }B^{\nu
}+(1/c)\varepsilon ^{\alpha \beta \mu \nu }v_{\mu }E_{\nu })e_{\beta } &=&0
\label{maeb}
\end{eqnarray}
where $E^{\alpha }$ and $B^{\alpha }$ are the basis components of the
electric and magnetic field 4-vectors $E^{a}$ and $B^{a}$ measured by a
family of observers moving with 4-velocity $v^{\alpha }$, and $\delta
_{\quad \mu \nu }^{\alpha \beta }=\delta _{\,\,\mu }^{\alpha }\delta
_{\,\,\nu }^{\beta }-\delta _{\,\,\nu }^{\alpha }\delta _{\,\,\mu }^{\beta
}. $ The equations (\ref{maeb}) correspond in the $E^{a},$ $B^{a}$ picture
to the equations (\ref{maco1}) in the $F^{ab}$ picture. From the relations (%
\ref{maeb}) we again find the covariant Maxwell equations for the basis
components (without the basis vectors $e_{\beta }$), which were already
presented in Refs. 2 and 3 
\begin{eqnarray}
\partial _{\alpha }(\delta _{\quad \mu \nu }^{\alpha \beta }v^{\mu }E^{\nu
}+c\varepsilon ^{\alpha \beta \mu \nu }B_{\mu }v_{\nu }) &=&-(j^{\beta
}/\varepsilon _{0})  \nonumber \\
\partial _{\alpha }(\delta _{\quad \mu \nu }^{\alpha \beta }v^{\mu }B^{\nu
}+(1/c)\varepsilon ^{\alpha \beta \mu \nu }v_{\mu }E_{\nu }) &=&0
\label{ma4}
\end{eqnarray}
(It has to be mentioned that the equations (\ref{ma4}) were also presented
in Ref. 23 but with $j^{\beta }=0.$ However in Ref. 23 the physical meaning
of $v^{\alpha }$ is unspecified - it is any unitary 4-vector. The reason for
such choice of $v^{\alpha }$ in Ref. 23 is that there $E^{\alpha }$ and $%
B^{\alpha }$ are introduced as the ''auxiliary fields,'' while $\mathbf{E}$
and $\mathbf{B}$ are considered as the physical fields. In our ''invariant''
approach with $E^{a}$ and $B^{a}$ the situation is just the opposite; $E^{a}$
and $B^{a}$ are the real physical fields, which are correctly defined and
measured in 4D spacetime, while the 3-vectors $\mathbf{E}$ and $\mathbf{B}$
are not correctly defined in 4D spacetime from the ''TT viewpoint.'') The
equations (\ref{ma4}) for basis components correspond to the covariant
Maxwell equations for basis components (\ref{maxco}), and the whole
discussion from the preceding section about the equations with $F^{ab},$
i.e., the CBGEs (\ref{maco1}) and the equations for basis components (\ref
{maxco}), can be easily translated to the equations with $E^{a}$ and $B^{a},$
(\ref{maeb}) and (\ref{ma4}). Instead of to work with $F^{ab}$- formulation,
(\ref{maco1}) and (\ref{maxco}), one can equivalently use the $E^{a},B^{a}$
formulation with (\ref{maeb}) and (\ref{ma4}). For the given sources $j^{a}$
one could solve these equations and find the general solutions for $E^{a}$
and $B^{a}.\bigskip \medskip $

\noindent \textbf{6.1. The Lorentz Force in Terms of }$E^{a}$\textbf{\ and }$%
B^{a}\medskip $

To complete the formulation of electrodynamics with $E^{a}$ and $B^{a}$ one
can write the expression for the Lorentz force in terms of $E^{a}$ and $%
B^{a} $ (see, e.g., Refs. 21 and 17) and also the equation of motion of a
charge $q $ moving in the electromagnetic field $E^{a}$ and $B^{a},$ Ref. 17$%
.$ The Lorentz force can be written in terms of $F^{ab}$ as $%
K^{a}=(q/c)F^{ab}u_{b}$ where $u^{b}$ is the 4-velocity of a charge $q$. It
has to be noted that usually the real physical meaning is not attributed to $%
F^{ab}$ but to the 3-vectors $\mathbf{E}$ and $\mathbf{B.}$ In the ''TT
relativity'' \emph{only the 4D tensor quantities }$F^{ab},$\emph{\ or }$%
E^{a} $\emph{\ and }$B^{a},$ \emph{do have well-defined physical meaning
both in the theory and in experiments. }Thence we express the Lorentz force
in terms of the 4-vectors $E^{a}$ and $B^{a}.$ In the general case of an
arbitrary spacetime and when $u^{a}$ is different from $v^{a}$ (the
4-velocity of an observer who measures $E^{a}$ and $B^{a}$), i.e. when the
charge and the observer have distinct world lines, $K^{a}$ can be written in
terms of $E^{a} $ and $B^{a}$ as a sum of the $v^{a}$ -orthogonal component, 
$K_{\perp }^{a}$, and $v^{a}$ -parallel component, $K_{\parallel }^{a}$, 
\begin{equation}
K^{a}=K_{\perp }^{a}+K_{\parallel }^{a}  \label{kei}
\end{equation}
$K_{\perp }^{a}$ is 
\begin{equation}
K_{\perp }^{a}=(q/c^{2})\left[ \left( -v^{b}u_{b}\right) E^{a}+c\widetilde{%
\varepsilon }^{a}\!_{bc}u^{b}B^{c}\right]  \label{kaokom}
\end{equation}
and $\widetilde{\varepsilon }_{abc}\equiv \varepsilon _{dabc}v^{d}$ is the
totally skew-symmetric Levi-Civita pseudotensor induced on the hypersurface
orthogonal to $v^{a}$, while 
\begin{equation}
K_{\parallel }^{a}=(q/c^{2})\left[ \left( E^{b}u_{b}\right) v^{a}\right]
\label{kapar}
\end{equation}
Speaking in terms of the prerelativistic notions one can say that in the
approach with the 4-vectors $E^{a}$ and $B^{a}$ $K_{\perp }^{a}$ (\ref
{kaokom}) plays the role of the usual Lorentz force lying on the 3D
hypersurface orthogonal to $v^{a}$, while $K_{\parallel }^{a}$ (\ref{kapar})
is related to the work done by the field on the charge. However \emph{in the
''TT relativity'' only both components together do have a physical meaning
and they define the Lorentz force both in the theory and in experiments. }

When \emph{the complete} $K^{a}$ ((\ref{kaokom}) \emph{and} (\ref{kapar}))
is known we can solve the equation of motion, Newton's second law, written
as tensor equation 
\begin{equation}
K^{a}=mu^{b}\nabla _{b}u^{a}  \label{eqmot}
\end{equation}
for the rhs of Eq. (\ref{eqmot}) see Ref. 11 Secs. 4.2 and 4.3. $\nabla _{b}$
is the derivative operator associated with $g_{ab}.$ (Note that in SR Wald$%
^{(11)}$ uses $\partial _{a},$ the ordinary derivative operator, instead of $%
\nabla _{a},$ compare his equations (4.3.2) in general relativity and
(4.2.26) in SR. However in the ''TT relativity'' formulation of SR one can
use different coordinatizations of an IFR. Thence, in general, the
derivatives of the nonconstant basis vectors must be also taken into
account, e.g., if one uses the ''e'' or ''r'' synchronization and polar or
spherical spatial coordinate basis. Therefore the use of the covariant
derivative $\nabla _{a}$ is necessary in SR as well.) The definition and the
measuring procedure for the 4-vectors $E^{a}$ and $B^{a}$ are determined by
the expression for $K^{a}$ and Newton's second law (\ref{eqmot}). The
comparison with the usual 3-vector form of the Lorentz force $\mathbf{F}=q%
\mathbf{E}+q(\mathbf{v}\times \mathbf{B})$ is considered in Ref. 17 and will
be reported in more detail elsewhere.\bigskip \medskip

\noindent \textbf{6.2. The Comparison of Maxwell's Equations with }$\mathbf{E%
}$\textbf{\ and }$\mathbf{B}$\textbf{\ and Those with }$E^{a}$\textbf{\ and }%
$B^{a}\medskip $

The comparison of this ''invariant'' approach with $E^{a}$ and $B^{a}$ and
the usual noncovariant approach with the 3-vectors $\mathbf{E}$ and $\mathbf{%
B}$ is possible in the ''e'' coordinatization, see also Ref. 17. If one
considers the ''e'' coordinatization and takes that in an IFR $S$ the
observers who measure the basis components $E^{\alpha }$ and $B^{\alpha }$
are at rest, i.e., $v^{\alpha }=(c,\mathbf{0})$, then $E^{0}=B^{0}=0$, and
one can derive from the covariant Maxwell equations (\ref{ma4}) for the
basis components $E^{\alpha }$ and $B^{\alpha }$ the Maxwell equations which
contain only the space parts $E^{i}$ and $B^{i}$ of $E^{\alpha }$ and $%
B^{\alpha }$, e.g., from the first covariant Maxwell equation in (\ref{ma4})
one easily finds $\partial _{i}E^{i}=j^{0}/\varepsilon _{0}c$. We see that
the Maxwell equations obtained in such a way from the Maxwell equations (\ref
{maeb}), or (\ref{ma4}), are of the same form as the usual Maxwell equations
with $\mathbf{E}$ and $\mathbf{B}$. From the above consideration one
concludes that all the results obtained in a given IFR $S$ from the usual
Maxwell equations with $\mathbf{E}$ and $\mathbf{B}$ remain valid in the
formulation with the 4-vectors $E^{a}$ and $B^{a}$ \emph{(in the ''e''
coordinatization), but only for the observers who measure the fields }$E^{a}$%
\emph{\ and }$B^{a}$\emph{\ and are at rest in the considered IFR.} Then for
such observers the components of $\mathbf{E}$ and $\mathbf{B}$, which are
not well defined quantities in the ''TT relativity,'' can be simply replaced
by the space components of the 4-vectors $E^{a}$ and $B^{a}$ (in the ''e''
coordinatization). It has to be noted that just such observers were usually
considered in the conventional formulation with the 3-vectors $\mathbf{E}$
and $\mathbf{B.}$ However, the observers who are at rest in some IFR $S$
cannot remain at rest in another IFR $S^{\prime }$ moving with $V^{\alpha }$
relative to $S$. Hence in $S^{\prime }$ this simple replacement does not
hold; in $S^{\prime }$ one cannot obtain the usual Maxwell equations with
the 3-vectors $\mathbf{E}^{\prime }$ and $\mathbf{B}^{\prime }$ (determined
by the AT (\ref{eitran})) from the transformed covariant Maxwell equations
with $E^{\alpha ^{\prime }}$ and $B^{\alpha ^{\prime }}$.

Some important experimental consequences of the ''TT relativity'' approach
have been derived in Ref. 3$.$ They are the existence of the spatial
components $E^{i}$ of $E^{a}$ outside a current-carrying conductor for the
observers (who measure $E^{a}$) at rest in the rest frame of the wire, and
the existence of opposite (invariant) charges on opposite sides of a square
loop with current, both when \emph{the loop is at rest} and when it is
moving.\bigskip \medskip

\noindent \textbf{6.3. The Covariant Majorana Form of Maxwell's Equations}
\medskip

We note that it is possible to write the Maxwell equations for $E^{a}$ and $%
B^{a}$ in another form, the covariant Majorana form, which is better suited
for the transition to the quantum physics. This can be realized by
introducing the covariant Majorana electromagnetic field 
\begin{equation}
\Psi ^{a}=E^{a}-icB^{a}  \label{psia}
\end{equation}
Then the covariant Majorana form of Maxwell's equations in the
coordinate-based geometric language and in the ''e'' coordinatization can be
determined from (\ref{maeb}) and it is 
\begin{equation}
((\gamma ^{\mu })^{\beta }\ _{\alpha }\partial _{\mu }\Psi ^{\alpha
})e_{\beta }=(-j^{\beta }/\varepsilon _{0})e_{\beta }  \label{Major}
\end{equation}
where the $\gamma $-matrices are 
\begin{equation}
(\gamma ^{\mu })^{\beta }\ _{\alpha }=\delta _{\rho \gamma }^{\mu \beta
}v^{\rho }g_{\alpha }^{\gamma }+i\varepsilon ^{\mu \beta }\ _{\alpha \gamma
}v^{\gamma }  \label{gama}
\end{equation}
>From (\ref{Major}) one finds the covariant Majorana form of Maxwell's
equations for the basis components $\Psi ^{\alpha }$ in the ''e''
coordinatization as 
\begin{equation}
(\gamma ^{\mu })^{\beta }\ _{\alpha }\partial _{\mu }\Psi ^{\alpha
}=-j^{\beta }/\varepsilon _{0}  \label{majo2}
\end{equation}

In the case that $j^{\beta }=0$ the equation (\ref{majo2}) becomes
Dirac-like relativistic wave equation for the free photon 
\begin{equation}
(\gamma ^{\mu })^{\beta }\ _{\alpha }\partial _{\mu }\Psi ^{\alpha }=0
\label{gam1}
\end{equation}
The similar equation was quoted and discussed in Ref. 23, but remember that
there is difference in the understanding of the physical meaning of $E^{a}$
and $B^{a},$ and thence of $\Psi ^{a},$ in Ref. 23 and in our approach (note
also that$^{(23)}$ exclusively deals with the basis components in the ''e''
coordinatization). From the Maxwell equations for the free electromagnetic
field (\ref{gam1}) one directly gets one-photon quantum equation
interpreting $\Psi ^{\alpha }$ as the one-photon wave function and
introducing the probability current and the continuity equation as in Ref.
23.

It can be seen from the recent literature that there is an increasing
interest in a photon wave function, one-photon quantum equation and in
photon localizability, see, e.g., Refs. 24-26 and references therein. The
common point for all these works is that they use the complex field $\mathbf{%
\Psi }$ as a linear combination of the electric and magnetic 3-vectors $%
\mathbf{E}$ and $\mathbf{B,}$ $\mathbf{\Psi =E}-ic\mathbf{B,}$ as in the
original Majorana idea. An unavoidable step in such treatments is that when
one wants to find the relation between the photon wave equation and the
usual Maxwell equations in the 3-vector form then one needs to get rid of
the Planck constant. In our invariant approach with $E^{a},$ $B^{a}$ and $%
\Psi ^{a}$ there is no need for such step. Furthermore it is considered as
one of the advantages of the use of the complex electromagnetic 3-vector $%
\mathbf{\Psi }$ that the energy density of the classical electromagnetic
field is equal to the square of the norm of that vector $\mathbf{\Psi .}$ $%
\mathbf{\Psi }^{*}\mathbf{\Psi }$ is thus interpreted as being directly
proportional to the probability density function for a photon. In the tensor
formulation with $E^{a},$ $B^{a}$ and $\Psi ^{a}$ one needs to use the
covariant expression for the energy-momentum density tensor $T^{ab}$ (the
equation (8) in Ref. 2 with $E^{\alpha }$ and $B^{\alpha }$, or the equation
(53) in Ref. 23 with $\Psi ^{\alpha };$ both expressions are actually the
basis components in the ''e'' coordinatization of the tensor $T^{ab}$)
instead of the usual expressions for the energy and momentum densities of
the electromagnetic field (with the 3-vectors $\mathbf{E,}$ $\mathbf{B}$ and 
$\mathbf{\Psi }$). We shall not further discuss the covariant Majorana
formulation since it will be reported elsewhere.\bigskip \medskip

\noindent \textbf{7. SUMMARY\ AND\ CONCLUSIONS\ }\bigskip

In this paper we have presented an invariant (true tensor) formulation, the
''TT relativity'' formulation, of SR. As stated in Sec. 2, the ''TT
relativity'' is the formulation of SR in which physical quantities in the 4D
spacetime are mathematically represented by true tensor fields (when no
basis has been introduced) that satisfy true tensor equations representing
physical laws. When some basis has been introduced the physical quantities
are described by the CBGQs, which contain both the components and the basis
one-forms and vectors of the chosen IFR, and which satisfy the CBGEs. It is
also shown in Sec. 2 that from the mathematical viewpoint the TT are the
isometries. These facts enable to treat different coordinatizations of an
IFR in the same manner. Two very different coordinatizations, the ''e'' and
''r'' coordinatizations, are exposed in Sec. 2 and exploited throughout the
paper. The ''TT relativity'' is compared with the usual covariant approach
to SR and with the usual ''AT relativity'' formulation, i.e., with the
original Einstein's formulation. In the usual covariant approach one deals
with the basis components of tensors and with the equations of physics
written out in the component form, and all is mainly done in the ''e''
coordinatization. In the ''AT relativity'' one does not deal with tensor
quantities but with quantities from ''3+1'' space and time, e.g., the
synchronously determined spatial lengths, or the temporal distances. The AT
connect such quantities and thus they refer exclusively to the component
form of tensor quantities and in that form they transform only \emph{some}
components of the whole tensor quantity.

The principal concept that makes distinction between the ''TT relativity''
formulation, the usual covariant formulation and the ''AT relativity''
formulation of SR is the concept of \emph{sameness} of a physical quantity
for different observers. In the ''TT relativity'' the same quantity for
different observers is the true tensor quantity, or equivalently the CBGQ,
only one quantity in 4D spacetime. The examples of such quantities are the
spacetime length $l$ (\ref{elspat}), Secs. 2 and 3, the distance 4-vector $%
l_{AB}^{a}$ (\ref{trucon}) and (\ref{trur}) and Fig. 1 in Sec. 3.1, and (\ref
{comu}) and (\ref{coer}) and Fig. 2 in Sec. 3.2. All these quantities are
mathematically equal $l_{AB}^{a}=l_{e}^{\mu }e_{\mu }=l_{e}^{\mu ^{\prime
}}e_{\mu ^{\prime }}=l_{r}^{\mu }r_{\mu }=l_{r}^{\mu ^{\prime }}r_{\mu
^{\prime }},$ (\ref{elt5}) in Sec. 2, and thus they are really the same
quantity for different observers. Note that these quantities are connected
by the TT, the Lorentz transformation $L^{a}{}_{b}$ (\ref{fah}), i.e., its
representations $L^{\mu ^{\prime }}{}_{\nu ,e}$ (\ref{lorus}) or $L^{\mu
^{\prime }}{}_{\nu ,r}$ (\ref{elr}), and the transformation $T_{\;\nu }^{\mu
}$ (\ref{lamb}) which connects different coordinatizations, Sec. 2. In the
usual covariant formulation of SR one considers that the basis components,
e.g., $l_{e}^{\mu }$ and $l_{e}^{\mu ^{\prime }}$, represent the same
quantity for different observers. These quantities, in fact, are not equal $%
l_{e}^{\mu }\neq l_{e}^{\mu ^{\prime }},$ but they only refer to the same
tensor quantity $l_{AB}^{a}$. If only one coordinatization is always used,
usually the ''e'' coordinatization, then the conventional covariant approach
can be applied. However the physics must not depend on the chosen
coordinatization, which means that the theory has to be formulated in the
manner that does not depend on the choice of some specific coordinatization.
The Einstein coordinatization is nothing more physical but the ''r''
coordinatization or any other permissible coordinatization. This requirement
is fulfilled in the ''TT relativity.'' The same thing can be also seen from
different relations in this paper, e.g., Eq. (\ref{fa2}) for the the
electromagnetic field tensor $F^{ab}$ Sec. 5.2. The analysis leading to Eqs.
(\ref{mxj}) and (\ref{mx2}) and these equations themselves clearly show why 
\emph{the ''TT relativity'' is an invariant }(in contrast to the usual
covariant) \emph{formulation of SR}, Sec. 5.3.

The same examples as in the ''TT relativity,'' Secs. 3.1 and 3.2, are
considered in the usual ''AT relativity'' in Secs. 4.1 and 4.2. In contrast
to the ''TT relativity'' the traditionally used ''AT relativity,'' considers
different \emph{spatial lengths }$l_{e}^{1},$\emph{\ }$L_{e}^{1^{\prime }},$%
\emph{\ }$l_{r}^{1},$\emph{\ }$L_{r}^{1^{\prime }},$ Fig. 3 (\emph{the
temporal distances }$l_{e}^{0},$\emph{\ }$L_{e}^{0^{\prime }},$\emph{\ }$%
l_{r}^{0},$\emph{\ }$L_{r}^{0^{\prime }},$\emph{\ }Fig. 4) as the same
quantity for different observers. The spatial lengths $L_{e}^{1^{\prime }}$\
and $L_{r}^{1^{\prime }}$ are connected with $l_{e}^{1},$\emph{\ }$%
l_{r}^{1}, $ i.e., with the rest length of the considered rod, by the
relations (\ref{apcon}) and (\ref{aper}) for the Lorentz ''contraction'' in
the ''e'' and ''r'' coordinatizations respectively. Similarly the temporal
distances $L_{e}^{0^{\prime }}$\ and $L_{r}^{0^{\prime }}$\emph{\ }are
connected with $l_{e}^{0},$\emph{\ }$l_{r}^{0},$\emph{\ }i.e., with the muon
lifetime at rest, by the relations (\ref{tidil}) and (\ref{tider}) for the
time ''dilatation'' in the ''e'' and ''r'' coordinatizations respectively.
The fact that the Lorentz ''contraction'' and the time ''dilatation''
connect different quantities in 4D spacetime proves that both
transformations are the AT.

In Sec. 5.1 we have presented the conventional derivation of the
transformations for the 3-vectors $\mathbf{E}$ and $\mathbf{B,}$ (\ref
{eitran}) and (\ref{eitr2}) by means of the identifications of the
components of $\mathbf{E}$ and $\mathbf{B}$ with \emph{some} of the basis
components of $F^{ab},$ the equations (\ref{eibi}) and (\ref{eib1}). But as
shown in Sec. 5.2 when only some of the basis components of a tensor
quantity are taken separately, as in the identifications (\ref{eibi}) and (%
\ref{eib1}), then they do not correspond to any definite physical quantity%
\emph{\ in 4D spacetime}. This can be interpreted saying that the
transformations (\ref{eitran}) and (\ref{eitr2}) connect different
quantities in 4D spacetime and thus that they are the AT. The same result is
obtained in Sec. 5.3 considering the derivation from$^{(18)}$ of (\ref{eitr2}%
) and (\ref{eitran}) which explicitly uses the Lorentz ''contraction'' (\ref
{apcon}). In Sec. 5.3 we have shown that the Maxwell equations in the
3-vector form (\ref{max}), or equivalently (\ref{eimax}) (always only in the
''e'' coordinatization), are not covariant (they change their form when
going from an IFR $S$ to another relatively moving IFR $S^{\prime },$ see (%
\ref{eima2})), and that they are not equivalent to the tensor Maxwell
equations (\ref{maxten}), or to the CBGEs (\ref{maco1}), (\ref{maxc2}), (\ref
{maxer}). Further we have examined the conventional, i.e., the ''AT
relativity'' definitions of charge by means of the volume integral of the
charge density and in terms of the Gauss law (for the 3-vector $\mathbf{E}$)
when written in the integral form Eq. (3) Sec. 5 in Ref. 18. Also Purcell's
formal statement of the relativistic invariance of charge is discussed. It
is shown that in the ''TT relativity'' these conventional definitions has to
be replaced by the tensor equations (\ref{charge}) and (\ref{gaus}).

In Sec. 6 we have introduced the 4-vectors $E^{a}$ and $B^{a}$ instead of
the usual 3-vectors $\mathbf{E}$ and $\mathbf{B}$ and we have formulated the
Maxwell equations as tensor equations with $E^{a}$ and $B^{a},$ i.e., as the
CBGEs (\ref{maeb}) and the equations for the basis components $E^{\alpha }$
and $B^{\alpha }$ (\ref{ma4}) (all in the ''e'' coordinatization). These
equations are completely equivalent to the usual covariant Maxwell equations
in the $F^{ab}$ formulation, (\ref{maco1}) and (\ref{maxco}). In Sec. 6.1 we
have presented the expressions for the Lorentz force in terms of the
4-vectors $E^{a}$ and $B^{a}.$ It has been explicitly shown in Sec. 6.2 that
all the results obtained in a given IFR $S$ from the usual Maxwell equations
with $\mathbf{E}$ and $\mathbf{B}$ remain valid in the formulation with the
4-vectors $E^{a}$ and $B^{a}$ (in the ''e'' coordinatization), but only for
the observers who measure the fields $E^{a}$\ and $B^{a}$\ and are at rest
in the considered IFR. Then in Sec. 6.3 we have constructed the Majorana
electromagnetic field four-vector $\Psi ^{a}$ (\ref{psia}) by means of
four-vectors $E^{a}$ and $B^{a}.$ The Maxwell equations in the covariant
Majorana formulation have been written in the coordinate-based geometric
language and in the ''e'' coordinatization as (\ref{Major}) and in the
component form as (\ref{majo2}). Using Majorana formulation we have found a
Dirac like relativistic wave equation for the free photon (\ref{gam1}). Our
next step will be the application of this covariant Majorana formulation to
the quantum electrodynamics.\bigskip \medskip

\newpage
\noindent \textbf{ACKNOWLEDGMENTS}\bigskip

It is a pleasure to acknowledge to Professor Larry Horwitz for inviting me
to the 2000 IARD conference, and also for his encouragement in preparing the
manuscript and for his useful suggestions and comments. Furthermore I am
grateful to Professor Alex Gersten for his hospitality and for helpful
discussions.\bigskip \medskip

\noindent \textbf{REFERENCES}\bigskip

\noindent \ 1. A. Einstein, \textit{Ann. Physik }\textbf{17,} 891 (1905),
tr. by W. Perrett and G.B.

\noindent \qquad Jeffery, in \textit{The principle of relativity} (Dover,
New York).

\noindent \ 2. T. Ivezi\'{c}, \textit{Found. Phys. Lett}. \textbf{12}, 105
(1999).

\noindent \ 3. T. Ivezi\'{c}, \textit{Found. Phys. Lett}. \textbf{12}, 507
(1999).

\noindent \ 4. F. Rohrlich, \textit{Nuovo Cimento B} \textbf{45}, 76 (1966).

\noindent \ 5. A. Gamba, \textit{Am. J. Phys}. \textbf{35}, 83 (1967).

\noindent \ 6. D.E. Fahnline, \textit{Am. J. Phys.} \textbf{50}, 818 (1982).

\noindent \ 7. T. Ivezi\'{c}, preprint Lanl Archives: physics/0007030.

\noindent \ 8. C. Leubner, K. Aufinger and P. Krumm, \textit{Eur. J. Phys.} 
\textbf{13,} 170 (1992).

\noindent \ 9. T. Ivezi\'{c}, preprint Lanl Archives: physics/0007031.

\noindent 10. J.D. Jackson, \textit{Classical Electrodynamics}, 2nd ed.
(Wiley, New York,

\noindent \qquad 1977).

\noindent 11. R.M. Wald, \textit{General relativity} (The University of
Chicago Press, Chicago,

\noindent \qquad 1984).

\noindent 12. B.F. Schutz, \textit{A first course in general relativity}
(Cambridge University

\noindent \qquad Press, Cambridge, 1985).

\noindent 13. C.W. Misner, K.S. Thorne and J.A. Wheeler, \textit{Gravitation}%
, (Freeman,

\noindent \qquad San Francisco, 1970).

\noindent 14. A.A. Logunov, \textit{Lectures in the theory of relativity and
gravity. A present-}

\noindent \qquad \textit{day analysis of the problem }(Nauka, Moskva, 1987)
(in Russian).

\noindent 15. R. Anderson, I Vetharaniam, G.E. Stedman, \textit{Phys. Rep.} 
\textbf{295}, 93 (1998).

\noindent 16. J. Norton, \textit{Found. Phys. }\textbf{19}, 1215 (1989).

\noindent 17. T. Ivezi\'{c}, preprint SCAN-9802018 (on the CERN server).

\noindent 18. E.M. Purcell, \textit{Electricity and magnetism}, 2nd.edn.
(McGraw-Hill, New

\noindent \qquad York, 1985).

\noindent 19. A. Einstein, \textit{Ann. Physik }\textbf{49,} 769 (1916), tr.
by W. Perrett and G.B.

\noindent \qquad Jeffery, in \textit{The principle of relativity }(Dover,
New York).

\noindent 20. H.W. Crater, \textit{Am. J. Phys.} \textbf{62}, 923 (1994).

\noindent 21. D.A. T. Vanzella, G.E.A. Matsas, H.W. Crater, \textit{Am. J.
Phys.} \textbf{64}, 1075

\noindent \qquad (1996).

\noindent 22. S. Sonego and M.A. Abramowicz, \textit{J. Math. Phys.} \textbf{%
39}, 3158 (1998).

\noindent 23. S. Esposito, \textit{Found. Phys.} \textbf{28}, 231 (1998).

\noindent 24. A. Gersten, \textit{Found. Phys. Lett}. \textbf{12}, 291
(1999).

\noindent 25. I. Bialynicki-Birula, \textit{Acta Phys. Pol.} A \textbf{86},
97 (1994); \textit{Phys. Rev. Lett.}

\noindent \qquad \textbf{80}, 5247 (1998).

\noindent 26. M. Hawton, \textit{Phys. Rev. A} \textbf{59}, 954 (1999); 
\textit{Phys. Rev. A} \textbf{59}, 3223 (1999)\newpage

\end{document}